\documentclass[aps,prr,twocolumn,groupedaddress,amsmath,epsfig,amssymb,eqsecnum]{revtex4}
\usepackage{makecell}
\usepackage{bbm}
\usepackage{mathrsfs}
\usepackage{subfigure}
\usepackage{multirow}
\usepackage{enumerate}
\usepackage{bm}
\usepackage{amsfonts}
\usepackage{color}
\usepackage{parskip}
\usepackage{graphicx}
\def\newblock{\hskip .11em plus .33em minus .07em}

\newcommand{\dbar}{d\hspace*{-0.08em}\bar{}\hspace*{0.1em}}
\usepackage{relsize}

\newcommand{\xv}{{\boldsymbol x}}

\newcommand{\be}{\begin{equation}}
\newcommand{\ee}{\end{equation}}
\newcommand{\ba}{\begin{eqnarray}}
\newcommand{\ea}{\end{eqnarray}}

\makeatletter
\newdimen\scalemath@axis
\newcommand*{\scalemath}[3]{%
  #1{%
    \mathpalette{\scalemath@aux{#2}}{#3}%
  }%
}
\newcommand*{\scalemath@aux}[3]{%
  \begingroup
    \everyvbox{}%
    \settoheight\scalemath@axis{$#2\vcenter{}$}%
    \raisebox{\scalemath@axis}{%
      \scalebox{#1}{%
        \raisebox{-\scalemath@axis}{%
          $\m@th#2#3$%
        }%
      }%
    }%
  \endgroup
}
\makeatother

\usepackage[doipre={doi:~}]{uri}

\usepackage[colorlinks=true,linkcolor=blue,anchorcolor=blue,citecolor=blue]{hyperref} 

\begin{document}
\title{Stochastic thermodynamics of Brownian motion in a flowing fluid}
\author{Jun Wu$^{1}$}
\author{Mingnan Ding$^{1}$}
\author{Xiangjun Xing$^{1,2,3}$}
\email{xxing@sjtu.edu.cn}
\affiliation{$^1$Wilczek Quantum Center, School of Physics and Astronomy, Shanghai Jiao Tong University, Shanghai, 200240 China\\
$^2$T.D. Lee Institute, Shanghai Jiao Tong University, Shanghai, 200240 China\\
$^3$Shanghai Research Center for Quantum Sciences, Shanghai 201315, China}

\date{\today}

\begin{abstract}
    We study stochastic thermodynamics of over-damped Brownian motion in a flowing fluid.   Unlike some previous works, we treat the effects of the flow field as a non-conservational driving force acting on the Brownian particle.  This allows us to apply the theoretical formalism developed in a recent work for general non-conservative Langevin dynamics. We define heat and work both at the trajectory level and at the ensemble level, and prove the second law of thermodynamics explicitly.  The entropy production (EP)  is decomposed into a housekeeping part and an excess part, both of which are non-negative at the ensemble level. Fluctuation theorems are derived for the housekeeping work, the excess work, and the total work, which are further verified using numerical simulations.  A comparison between our theory and an earlier theory by Speck et. al. is also carried out. 
   
\end{abstract}
\maketitle
\clearpage

\section{Introduction}
The theory of Brownian motion~\cite{einstein_1905_motion,zwanzig_2001_nonequilibrium} is not Galilean invariant, even though the underlying microscopic Newtonian dynamics does have this symmetry.  The reason for the lack of Galilean symmetry is obvious: the ambient fluid is macroscopically at rest only in one particular inertial frame. It is only in this frame that the effects of fluid can be modeled as friction and random force as in the classical theory of Brownian dynamics. If the fluid is in global motion with a uniform velocity $\bm v$,  one can transform to the co-moving frame (where the fluid is at rest) and apply the usual theory of Brownian motion. Translating back into the lab frame, the friction force becomes $- \gamma (\dot \xv - \bm v)$, which is proportional to the velocity relative to the fluid, whereas the random force remains the same.  Now consider a fluid that is shearing or compressing, with a position (and possibly time) dependent velocity $\bm v(\xv, t)$.  The above chain of argument is not as compelling, since the co-moving frame is constantly deforming and therefore is not an inertial frame.   Nonetheless, as long as the flow field is small, it is reasonable to assume that the friction force is approximately $-\gamma (\dot \xv - \bm v(\xv,t) )$. 

Stochastic thermodynamics~\cite{gallavotti_1995_dynamical,searles_1999_fluctuation,seifert_stochastic_2012,peliti2021stochastic,van2013stochastic,jarzynski_2011_equalities,crooks_1999_entropya} fuses stochastic dynamics with thermodynamics to form a unified framework for non-equilibrium statistical mechanics.  In the standard theory of stochastic dynamics, the environment is usually assumed to be an equilibrium fluid at rest.   In a pioneering work~\cite{speck_2008_role}, Speck {\it et. al.} applied the above idea of Galilean transform to study stochastic thermodynamics of over-damped Brownian motion in a moving fluid. In the co-moving frame, the heat is defined as the work done by the friction and the random force, as in the standard theory of stochastic thermodynamics. Further defining the external potential as the fluctuating internal energy of the Brownian particle, they sketched a framework of stochastic thermodynamics for over-damped Brownian motion in moving fluid.  This work was followed by several later  works~\cite{li_2021_galilean,cairoli_2018_weak,gerloff_2018_stochastic,ghosal_2016_polymer,pagare_2019_stochastic,ghosal_2019_fluctuation}, which carried out more detailed analyses and simulations of fluctuation theorems in shearing fluid.  Three peculiar features are coming out of this theory.  (For details, see Sec.~\ref{sec:difference} of the present work.)  Firstly it leads to fluctuation theorems for the integrated work only if the non-equilibrium processes start from equilibrium states where the flow field is completely turned off.  The theory is therefore inapplicable to processes happening in steady flow. Secondly, the entropy production (EP)  is positive only for incompressible flow.   Finally, the EP in this theory cannot be decomposed into two positive parts. Hence no separate fluctuation theorem can be established for the housekeeping part and the excess part of the total EP.  This is at odds with the basic structure of stochastic thermodynamics for systems lacking instantaneous detailed balance, as established by Jarzynski and Esposito, van den Brock et. al.~\cite{chernyak_2006_pathintegral,esposito_2010_three, esposito_2010_threea,vandenbroeck_2010_three,ding_unified_2022}. 

It is well known that heat in stochastic thermodynamics is related to the time reversal of non-equilibrium processes through the condition of {\em local detailed balance}.   It turns out that in the theory of Ref.~\cite{speck_2008_role}, time reversal means reversal of both the time axis and the flow field.   Since the environment, i.e. the shearing fluid has a well-defined temperature, the heat is further related to the environmental entropy change via $\Delta S^{\rm env} = - \beta Q$.  It is important to note, however, that the entropy change calculated this way is a microscopic quantity, whereas the true entropy change of the environment is extensive in the size of the shear fluid.  Hence the EP calculated in the theory of Ref.~\cite{speck_2008_role} can only be a tiny part of the physical EP in the joint system of Brownian particle and shearing fluid.  This subtlety is shared by all models of stochastic thermodynamics embedded in dissipative backgrounds, such  as Brownian motion in temperature gradients~\cite{ding_2024_stochastic}.  With this subtlety carefully remembered, the fluctuation theorems, when formulated in terms of integrated work, are nonetheless valuable tools for understanding of the statistical properties of non-equilibrium processes. 

In this work, we shall try an alternative approach to the problem.  Instead of transforming to the co-moving, we shall stay in the lab frame and treat the effects of the flow field as a non-conservative force acting on the Brownian particle.  This allows us to apply the general framework of stochastic thermodynamics developed in Ref.~\cite{ding_unified_2022} for non-conservative Langevin systems.  In our theory, the time reversal of process means the reversal of only the time-axis, but not of the flow field.  Consequently, the heat defined in our theory is inequivalent to that defined in Ref.~\cite{speck_2008_role}.  As discussed in great detail in Ref.~\cite{ding_unified_2022}, in the absence of instantaneous detailed balance, there are also ambiguities in the definition of system energy.  Different definitions of energy lead to different (but equivalent) formulations of stochastic thermodynamics.  The situation is not unlike the gauge redundancy in electromagnetism.  We shall adopt the so-called Gibbs gauge where the instantaneous non-equilibrium steady state (NESS)  has the form of Gibbs-Boltzmann distribution, which leads to great simplification of the theoretical formalism.  

The key results of the present work can be summarized as follows: (i) The  EP at the ensemble level that emerges from our theory is positive definite for arbitrary flow field.  (ii)  The  EP  can be decomposed into a housekeeping part and an excess part, both of which are positive.  (iii) At the trajectory level, both the housekeeping work and the excess work obey a fluctuation theorem. (iv) These fluctuation theorems are applicable for arbitrary processes starting from arbitrary non-equilibrium NESSs, including equilibrium states as special cases.  Overall, therefore, the present theory has a wider range of applicability than that of Ref.~\cite{speck_2008_role}.  

The remaining of this work is organized as follows.  In Sec.~\ref{sec:Brownian-dyn} we formulate the Langevin equation for Brownian motion in a following fluid. In particular, in Sec.~\ref{sec:adjoint} we discuss the adjoint Brownian dynamics, in Sec.~\ref{sec:harmonic} we perturbatively calculate the Gibbs gauge representation. 
 In Sec.~\ref{sec:ST} we develop the theory of stochastic thermodynamics and derive fluctuation theorems for the housekeeping EP, the excess EP, and the total EP.  In Sec.~\ref{sec:difference}, we discuss the differences between our theory and the theory of Ref.~\cite{speck_2008_role}.  In Sec.~\ref{sec:simulation} we present numerical verifications of all fluctuation theorems.  Finally in Sec.~\ref{sec:conclusion} we draw concluding remarks and project future research directions.

\section{Brownian dynamics in a flow }
\label{sec:Brownian-dyn}
\subsection{Langevin equation}
For simplicity, we examine two-dimensional Brownian dynamics in a fluid with time-independent flow.  Generalization to three-dimensional time-dependent flow is straightforward.    The velocity field of the fluid is  
\begin{equation}
    {\bm v}({\bm x}) =v_x (\bm x) \, \hat{\bm e}_x + v_y(\bm x) \, \hat{\bm e}_y,
\end{equation}
where $\hat{\bm e}_x,\hat{\bm e}_y $ are respectively the unit vectors along $x$ and $y$ directions, and $\bm x = x \, \hat{\bm e}_x + y \, \hat{\bm e}_y$.    Assuming that the Brownian particle is further confined by an external potential $V(\bm x)$, its motion can be described by the following over-damped  Ito-Langevin equations:
\be
\begin{split}
    - \gamma (dx-v_x dt)-\partial_x V dt + \sqrt{2\gamma T} \, dW_x  & = 0, \\
  -  \gamma (dy-v_y dt ) -\partial_y V dt + \sqrt{2\gamma T} \, dW_y  & = 0,
\end{split}
\label{langevin-1}
\ee
where $\gamma$ is the friction constant, $T$ is the temperature, and $dW_{x}, dW_{y}$ are the standard Wiener noises, which have the following basic properties:
\begin{subequations}
\ba
\langle dW_i \rangle &=& 0, \\
 \langle dW_i dW_j \rangle &=& dt \, \delta^{ij}. 
\ea
\label{Wiener-properties}
\end{subequations}
Note that the first term in each of Eqs.~(\ref{langevin-1}) is the friction force multiplied by $dt$.   

Equations (\ref{langevin-1})  can be rewritten into:
\ba
    dx^i +  \frac{T}{\gamma}  ( \partial_i U^0 - \varphi_i^0 )  dt 
   = \sqrt{\frac{2 T}{\gamma}}  dW_i, 
\label{langevin-2}
\ea
where $U^0, {\bm \varphi}^0$ are given respectively by
\begin{subequations}
\ba
U^0(\bm x) &=& \beta \, V(\bm x) + C^0, 
\label{U-varphi-def-U}\\
{\bm \varphi}^0 (\bm x) &=& \beta \gamma\, {\bm v}(\bm x)
=  \varphi^0_i (\bm x) \hat {\bm e}_i, \quad
\label{U-varphi-def-varphi}
\ea
\label{U-varphi-def}
\end{subequations}
where $C^0$ is an irrelevant normalization constant.   We do not need to distinguish superscripts from subscripts because we will only use Cartesian coordinate systems.  Equation (\ref{langevin-2}) is a special case of the following covariant nonlinear Ito-Langevin equation with non-conservative forces~\cite{ding_unified_2022} (with all repeated indices summed over):
\begin{equation}
    dx^i +  L^{ij} ( \partial_j U^0 - \varphi_j^0 )  dt -\partial_j L^{ij} dt
     = b^{i\alpha}dW_\alpha(t),
   \label{Langevin-4}  
\end{equation}
where all variables are even under time reversal, and the  $2\! \times\! 2$ matrices $L^{ij}$ and $b^{i \alpha}$ are given respectively by
\ba
{\bm b} &=& 
\sqrt{\frac{2 T}{\gamma}} 
\begin{pmatrix}
1
&0\\
0 & 1
\end{pmatrix},\\
{\bm L} &=& {\bm L}^T =
\frac{T}{\gamma}
 \begin{pmatrix}
1
&0\\
0 &1
\end{pmatrix}
= \frac{1}{2} {\bm b} {\bm b}^T.
\ea
Since both $T$ and $\gamma$ are constants, $\partial_j L^{ij}$ in Eq.~(\ref{Langevin-4}) vanishes identically.  We shall call $U^0$ and $\varphi_i^0$ respectively {\em the generalized potential} and {\em the non-conservative force}~\footnote{Strictly speaking, the non-conservative force is defined as $T \bm \varphi$ in Ref.~\cite{ding_unified_2022}.}. 


The Langevin equation (\ref{langevin-2}) is equivalent to the following covariant Fokker-Planck equation (FPE):
\ba
    \partial_t p -   \partial_i \frac{T}{\gamma} 
    (\partial_i + \partial_i U^0 
    - \varphi_i^0 ) p = 0, 
    \label{FPE-1}
\label{equ::FP equation} 
\ea
which can also be written in the form of:
\ba
\partial_t p + \partial_k j_k = 0,
\ea
where $j_k$ is the probability current:
\ba
j_i =  - \frac{T}{\gamma} (\partial_i + (\partial_i U^0) - \varphi_i^0 ) p.
\label{current-def}
\ea

It is easy to see that the following  transformation:
\begin{subequations}
\ba
U^0 &\rightarrow& U = U^0 + \psi, \\
\varphi_i^0 &\rightarrow&\varphi_i 
= \varphi_i^0  + \partial_i \psi,
\ea
\label{gauge-transform}
\label{Gibbs-gauge-1}
\end{subequations}
leaves the combination $\partial_i U^0 - \varphi_i^0$ invariant, and hence also 
leaves the Langevin equation (\ref{Langevin-4}) and the Fokker-Planck equation (\ref{equ::FP equation}) as well as the probability current (\ref{current-def}) invariant.  Inspecting Eqs.~(\ref{U-varphi-def-U}) and (\ref{U-varphi-def-varphi}), we see that the transform (\ref{gauge-transform}) may be understood as a simultaneous change of the external potential and the flow field that preserves the Brownian motion.  We shall call it {\em a gauge transformation}.    A particular decomposition of the combination $\partial_i U - \varphi_i$ into $\partial_i U$ and $\varphi_i $ shall then be called a {\em gauge}.  

The most convenient gauge is the {\em Gibbs gauge}~\cite{ding_unified_2022}, where $U$ is related to the NESS via
\ba
p^{\rm ss}(\bm x) = e^{- U(\bm x)}. 
\label{p^ss-def-1}
\ea
Substituting this back into Eq.~(\ref{current-def}), we find the NESS probability current is then given by
\ba
j_i^{\rm ss}(\bm x)  =  \frac{T}{\gamma} e^{-U(\bm x) } \varphi_i(\bm x) .
\label{j^ss-def-1}
\ea
which is proportional to $\varphi_i$.  The fact that $\varphi_i$ is non-vanishing characterizes the non-equilibrium nature of the NESS.  Substituting Eq.~(\ref{j^ss-def-1}) into the steady state FPE $\nabla \cdot {\bm j}^{\rm ss}  = 0$, we obtain the {\em Gibbs gauge condition}:
\begin{gather}
\partial_i \varphi_i - (\partial_i U) \varphi_i  = 0,
     \label{Gibbs-gauge-cond}
\end{gather}
which,  using Eq.~(\ref{Gibbs-gauge-1}), can be further rewritten into:
\ba
\partial_i ( \varphi_i^0  + \partial_i \psi) 
- ( \varphi_i^0  + \partial_i \psi) \partial_i  (U^0 + \psi)  = 0. 
     \label{Gibbs-gauge-cond-1}
\ea
In Sec.~\ref{sec:harmonic}, we solve this nonlinear differential equation for the case of simple shear flow and harmonic confining potential, and use Eqs.~(\ref{Gibbs-gauge-1}) to determine $U, \varphi_i$ for the Gibbs gauge.

In the Gibbs gauge, the Langevin equation and the FPE are given by
 \begin{subequations}
\ba
    dx^i +  \frac{T}{\gamma}  ( \partial_i U - \varphi_i )\,  dt 
   &=& \sqrt{\frac{2 T}{\gamma}}  dW_i, 
   \label{Langevin-Gibbs}  \quad\quad \\
     \partial_t p - \frac{T}{\gamma}  \partial_i 
    (\partial_i + \partial_i U 
    - \varphi_i ) \, p     &=& 0.   
      \label{FPE-Gibbs} 
\ea
   \label{Langevin-dynamics-1}
   \end{subequations} 
Equation (\ref{Langevin-Gibbs}) will be called {\em the Gibbs representation} of the Langevin dynamics, whereas Eqs.~(\ref{langevin-1}) and (\ref{langevin-2}) will be called {\em the natural representation} of the Langevin dynamics.

\subsection{Adjoint Langevin dynamics}
\label{sec:adjoint}
We now define {\em the adjoint Langevin dynamics}, which is related to the original dynamics (\ref{Langevin-dynamics-1}) via the following transform in the Gibbs gauge:
\ba
 U^{\rm Ad} = U, \quad \varphi_i^{\rm Ad}  = - \varphi_i. 
 \label{U-phi-Ad-1}
\ea 
Using Eqs.~(\ref{p^ss-def-1}) and (\ref{j^ss-def-1}), we see that the adjoint process has the same NESS pdf and opposite NESS probability current as the original process:
\ba
p^{\rm Ad, ss}(\bm x) &=& e^{- U(\bm x)}
= p^{\rm ss}(\bm x) .
\label{p^ss-def-1-ad}\\
j_i^{\rm Ad, ss}(\bm x)  &=& - \frac{T}{\gamma} e^{-U(\bm x) } \varphi_i(\bm x) 
= - j_i^{\rm ss}(\bm x)  .
\label{j^ss-def-1-ad}
\ea


 Just as the original dynamics, the adjoint dynamics can also be realized by many different combinations of flow field and confining potential, each characterized by a pair $\{ U^{0, \rm Ad}, \varphi_i^{0, \rm Ad} \}$ that is related to $\{U^{\rm Ad}, \bm \varphi^{\rm Ad} \} $ via a gauge transformation:
\begin{subequations}
\ba
U^{\rm 0, Ad} &\rightarrow& U^{\rm Ad} 
	= U^{\rm 0, Ad} + \psi^{\rm Ad}, \\
\varphi_i^{\rm 0, Ad} &\rightarrow& \varphi_i^{\rm Ad} 
	= \varphi_i^{\rm 0, Ad}  + \partial_i \psi^{\rm Ad},
\ea
\label{gauge-transform-Ad}
\label{Gibbs-gauge-1-Ad}
\end{subequations}
which is the counterpart of Eqs.~(\ref{Gibbs-gauge-1}).  The gauge function $\psi^{\rm Ad}$ is arbitrary.  The most convenient choice is:
\ba
\psi^{\rm Ad} &=& - \psi.
\ea
Substituting this back into Eqs.~(\ref{Gibbs-gauge-1-Ad}), we may express $\{ U^{0, \rm Ad}, \varphi_i^{0, \rm Ad} \}$ in terms of $\{U^{\rm Ad}, \bm \varphi^{\rm Ad},\psi \} $.  Combining these results with Eqs.~(\ref{U-phi-Ad-1}), we obtain:
 \begin{subequations}
\ba
U^{0, \rm Ad} &=& U^0 + 2 \psi;\\
\varphi_i^{0, \rm Ad} &=& - \varphi_i^0. 
\ea
\label{phi-U-psi-Ad}
\end{subequations}
Substituting these into Eqs.~(\ref{U-varphi-def}), we find the confining potential and the flow field for the adjoint process:
\begin{subequations}
\label{V-v-Ad}
\ba
V^{\rm Ad}(\xv) &=& T\, (U^0 + 2\,\psi - C^{0})
\nonumber\\
&=& V (\xv) + 2\,T \, \psi(\xv), \\
\bm v^{\rm Ad} (\xv) &=& - \frac{T}{\gamma}  \bm \varphi 
= - \bm v(\xv).
\ea
\end{subequations}
Hence the flow field of the adjoint dynamics is the opposite of that of the original process.  

The Gibbs representation of the adjoint Langevin dynamics can be obtained by using Eq.~(\ref{U-phi-Ad-1}) in Eq.~(\ref{Langevin-Gibbs}), whereas the natural representation of the adjoint Langevin dynamics can be obtained by using Eqs.~(\ref{phi-U-psi-Ad}) in Eq.~(\ref{langevin-2}), or, equivalently, by using Eqs.~(\ref{V-v-Ad}) in Eqs.~(\ref{langevin-1}). 



\subsection{Harmonic potential and simple shear flow}
\label{sec:harmonic}

\begin{figure}[t!]
  \centering
  \includegraphics[width=1.9in]{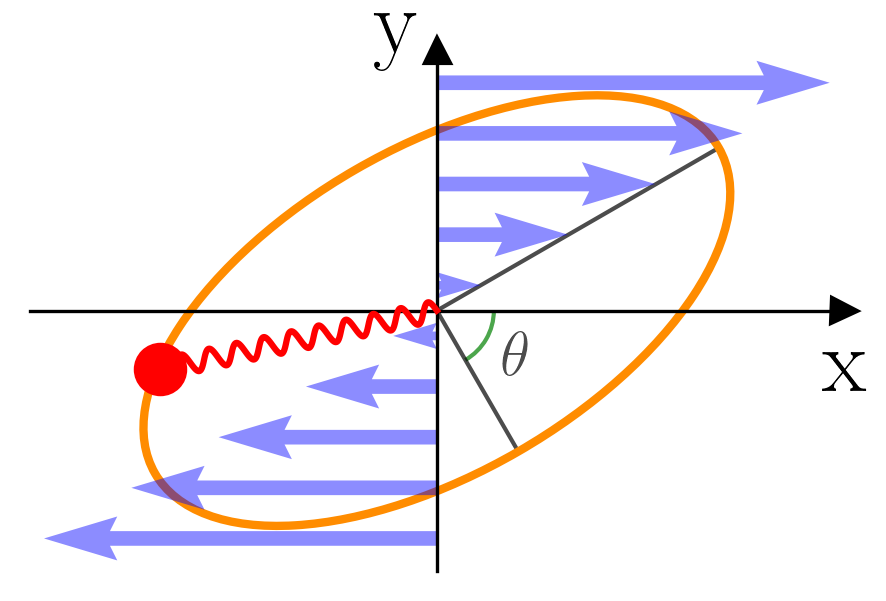}
    \vspace{-3mm}
  \caption{Schematics: red disk is the Brownian particle, blue arrows represent the simple shear flow, whereas the red wiggly line represents the harmonic potential. The orange ellipse is a contour line of the NESS pdf.  $\theta$ is the angle between the major axis and the $y$ axis, or equivalently the angle between the minor axis and the $x$ axis.}
  \label{fig::sketch_ellipse}
\end{figure}

For numerical simulations (to be detailed in Sec.~\ref{sec:simulation}), we shall only consider a harmonic confining potential and a simple shear flow:
\begin{subequations}
\ba
V (\bm x) &=& \frac{K}{2} (\bm x - \bm x_0)^2, 
\label{V-v-1-V}
\\
{\bm v}(\bm x) &=& {y \, \zeta}\,  \hat{\bm e}_x. 
\ea
\label{V-v-1}
\end{subequations}
Using Eqs.~(\ref{U-varphi-def}) we find 
\begin{subequations}
\ba
U^0 &=& \frac{\beta K}{2}  (\bm x - \bm x_0)^2 + C^0,
\label{U0-phi0-U0}\\
{\bm \varphi}^0 (\bm x) &=& 
{\beta \gamma} \zeta \, y\, \hat{\bm e}_x,
\ea
\label{U0-varphi0-1}
\end{subequations}
Note that $\zeta$ has the dimension of inverse time, and $\gamma \, \zeta$ is even under time reversal. The dimension of $\bm \varphi$ is inverse of length, and hence also even under time reversal.   
  The shear flow and the confining potential are illustrated in Fig.~\ref{fig::sketch_ellipse}, together with a contour line of the NESS pdf. 

We define a dimensionless parameter $\epsilon \equiv {\gamma \, \zeta}/{K}$,  
which characterizes the relative importance of the flow field compared with the confining potential.  For colloidal particles in shearing fluid under typical experimental conditions, this parameter is expected to be much less than the unity, hence we expect that the flow field only leads to a small perturbation of the equilibrium distribution of the Brownian particle.  Therefore we may solve Eq.~(\ref{Gibbs-gauge-cond-1}) by expanding $\psi$ in terms of $\epsilon$, and subsequently use Eqs.~(\ref{Gibbs-gauge-1}) to find $U$ and $\bm \varphi$.  Since the calculation is rather straightforward, we skip all details and directly present the second order results:
\begin{subequations}
 \label{Gibbs-parameters}
\ba
\psi  &=& \frac{\beta K \epsilon }{2 }
(  -   xy - x y_0 + x_0 y  )
\nonumber\\
&+& \frac{\beta K \epsilon^2 }{8 }
 ( - x^2 + y^2 + 2 x_0 x  + 2 y_0 y )
 , \label{equ::2nd order psi}  \\
U &=&C + U^0 +   \psi ,\\
\bm \varphi &=& 
 \frac{\beta K \epsilon}{2}
\left[ 
(y - y_0) \hat{\bm e}_x
 + ( - x + x_0) \,\hat{\bm e}_y
 \right] 
 \nonumber\\
&+ &  \frac{\beta K \epsilon^2}{4}
\left[ 
(-x + x_0) \hat{\bm e}_x
 + (y + y_0) \, \hat{\bm e}_y
 \right], \quad 
 \ea
 \end{subequations}
 where the constant $C$ is such that Eq.~(\ref{p^ss-def-1}) is normalized.  We shall not need the concrete expression for $C$. The reader may verify directly that Eqs.~(\ref{equ::2nd order psi}) does satisfy the Gibbs gauge condition (\ref{Gibbs-gauge-cond}) up to $o(\epsilon^2)$.  
 
If $\epsilon$ is not small,  Eqs.~(\ref{Gibbs-parameters}) are not good approximations. Nonetheless, it is easy to see from Eq.~(\ref{Gibbs-gauge-cond-1})   that, due to the quadratic nature of $U^0$ and the linear nature of $\varphi^0_i$, $\psi$ is quadratic in $\xv$.  Consequently, $U$ is also quadratic in $\xv$, and $\bm \varphi$ is linear in $\xv$. Contour lines of $U$ are therefore all ellipses, one of them being illustrated in Fig.~\ref{fig::sketch_ellipse}.  We can therefore set 
\ba
  U = A x^2 + B x y + C y^2 + D x + E y + F,
  \label{equ::U in ABC form-0}
\ea
and numerically find all coefficients $A,B, C, D, E$. $F$ is determined by the condition of normalization. The numerical method is explained in App.~\ref{sec:Num-Adj}.


\section{Stochastic thermodynamics}
\label{sec:ST}
In Ref.~\cite{ding_unified_2022}, we developed a unified theory of stochastic thermodynamics for Langevin systems driven by non-conservative forces and coupled to a single heat bath with temperature $T$.  The Langevin dynamics Eq.~(\ref{Langevin-4}) is a special case of this unified theory, with all variables and control parameters being even under time reversal, and the kinetic matrix $\bm L$ being symmetric and constant. We shall therefore follow the procedure developed in Ref.~\cite{ding_unified_2022} (especially Sec. VI, which treats symmetric models) 
to develop the theory of stochastic thermodynamics.

It is important to emphasize that by formulating the Langevin equation into Eq.~(\ref{Langevin-dynamics-1}), we are taking the viewpoint that the effects of the flow field is treated as a non-conservative force field.  The dissipation caused by the shearing fluid, which is extensive in the size of the fluid, is not a concern to us, since we are only interested in the dynamics of the Brownian particle.  

In this work, we shall assume that the flow field is fixed and consider non-equilibrium processes where the force constant $K$ and the equilibrium position $\xv_0$ of the confining potential (\ref{V-v-1-V}) are systematically varied.  For simplicity, we introduce the notations $\lambda = \{ K, \xv_0 \}$, and $\lambda_t = \{ K(t), \xv_0(t) \}$.  It then follows that $V, U^0$ as well as  $U, \bm \varphi, \psi$ all depend parametrically on $\lambda_t$.  We will therefore use the notations $V(\xv; \lambda_t), U^0(\xv; \lambda_t)$ etc..  In principle, the theory we develop is also applicable to processes where the flow field is also systematically varied.   Experimentally, however, it is much more difficult to vary the flow field in a precisely controlled way. 

\subsection{Work, heat and  EP }
\label{sec:W-Q-Sigma}

We define the fluctuating internal energy as 
\ba
H(\bm x; \lambda) \equiv T  U(\xv; \lambda)  = T (U^0 + \psi),
\label{H-def-1}
\ea
where $U(\xv; \lambda)$ is the generalized potential in Gibbs gauge, to be found by solving Eq.~(\ref{Gibbs-gauge-cond}).  The NESS distribution Eq.~(\ref{p^ss-def-1}) can then be rewritten as 
\ba
p^{\rm ss}(\bm x; \lambda) 
= e^{ - \beta H (\bm x; \lambda)} 
= e^{-U(\bm x; \lambda)}.
\label{p^ss-1}
\ea
To realize such a NESS, one only needs to hold the shear flow and the control parameter $\lambda$ fixed for a sufficiently long period of time.  {\em The equilibrium free energy} $F(\lambda)$ is defined as
\ba
F(\lambda) \equiv - T \, \log \int_{\bm x} e^{-\beta H} = - T \, \log 1 = 0, 
\ea
where in the second step we used the fact that $p^{\rm ss}(\bm x)$ is properly normalized.  Hence  our definition (\ref{H-def-1}) of energy guarantees that the equilibrium free energy vanishes identically.  This leads to certain simplification of the fluctuation theorems, as we shall see below. 

We define differential heat at the trajectory level as 
\ba
\dbar {\mathscr Q} &\equiv& d_{\bm x} H
 - T\, \bm \varphi \circ d \xv 
\nonumber\\
&=& T\, ( d_{\bm x} U  -  \bm \varphi \circ d \xv ),
\label{dQ-traj-1}
\ea
where $d_{\bm x} H$ means differential of $H$  due to variation of $\bm x$, and $\circ$ is the stochastic product in Stratonovich's sense:
 \ba
  \bm \varphi \circ d \xv \equiv  \varphi_i(\xv + d \xv/2) \,d x^i.
 \ea
 The differential work at the trajectory level is defined as 
\ba
\dbar {\mathscr W} &\equiv& d_\lambda H
 + T \, \bm \varphi \circ d \xv.
\nonumber\\
&=& T\, ( d_\lambda U  + \bm \varphi \circ d \xv) ,
\label{equ::W-def}
\ea
where $d_\lambda H$ is the differential of $H$ due to variation of $\lambda$.  


The above defined heat and work can be decomposed into a housekeeping part and an excess part:
\ba
\dbar {\mathscr Q} &=& \dbar {\mathscr Q}^{\rm hk} + \dbar {\mathscr Q}^{\rm ex},\\
\dbar {\mathscr W} &=& \dbar {\mathscr W}^{\rm hk} + \dbar {\mathscr W}^{\rm ex}, 
\ea
where $\dbar {\mathscr Q}^{\rm hk},\dbar {\mathscr W}^{\rm hk}$ are respectively housekeeping heat and housekeeping work, whereas $ \dbar {\mathscr Q}^{\rm ex}, \dbar {\mathscr W}^{\rm ex}$ are respectively excess heat and excess work,  defined as
\begin{subequations}
\label{Q-W-hk-ex-def}
\ba
\dbar {\mathscr Q}^{\rm hk} &\equiv& 
 - T\, \bm \varphi \circ d \xv  ,
  \label{dQ-hk-def} \\
\dbar {\mathscr Q}^{\rm ex} &\equiv& 
 T\, d_{\bm x} U , 
 \label{dQ-ex-def}\\
\dbar {\mathscr W}^{\rm hk} &\equiv& 
T\, \bm \varphi \circ d \xv
= - \dbar {\mathscr Q}^{\rm ex},
 \label{W-hk-def}\\
\dbar {\mathscr W}^{\rm ex} &\equiv& 
 T\, d_\lambda U .
  \label{W-ex-def}
\ea
\end{subequations}
The first law of thermodynamics at the trajectory level is given by either of the following two forms:
\ba
\begin{split}
d H &= \dbar {\mathscr Q} +  \dbar {\mathscr W} 
=  \dbar {\mathscr Q}^{\rm ex}  + \dbar {\mathscr W}^{\rm ex}.
\end{split}
\ea
Note that the housekeeping heat and housekeeping work exactly cancel each other. 

Heat and work at the ensemble level can be obtained by averaging the corresponding quantities at the trajectory level over both noises and the pdf of $\xv$.  To obtain a well-defined continuum limit, these differential quantities at the ensemble level must be computed up to the first order in $dt$.  It is important to remember that the Wiener noises are square root of $dt$, and hence according to Eq.~(\ref{Langevin-Gibbs}), $d \xv$ contains parts scaling with $\sqrt{dt}$.  Consequently, we need to expand these differential quantities up to the second order of $d\xv$, to keep all terms linear in $dt$. 

As an example, let us compute the excess heat at the ensemble level:
\ba
\dbar Q^{\rm ex} &=& \langle \!\langle \dbar {\mathscr Q}^{\rm ex} \rangle \! \rangle,
\ea
where $ \langle \!\langle \, \cdot \, \rangle \! \rangle$ means double average over Wiener noises and over probability distribution of $\bm x$.   First we use Eq.~(\ref{dQ-ex-def}) to expand $\dbar {\mathscr Q}^{\rm ex} $ up to the second order in $d\xv$:
\ba
\dbar {\mathscr Q}^{\rm ex}  
= T \partial_i U d x^i
+ \frac{T}{2} \partial_i \partial_j U dx^i dx^j,
\ea
where all products are in Ito's sense.  We now use the Langevin equation (\ref{Langevin-Gibbs}) to express $d\xv$ in terms of $dt$ and $d \bm W$.  All terms in the form of $dt^2$ and $dt dW_i$ can be neglected, since they are higher order than $dt$.  Then we average over the Wiener noises $d \bm W$.
Finally we multiply the result by the pdf $p(\xv; t)$ and integrate over $\xv$, and obtain the differential excess heat at the ensemble level:
\ba
\dbar Q^{\rm ex} =  - \frac{T^2\, dt}{\gamma} 
\left\langle  \partial_i U (\partial_i U - \varphi_i )
- \partial_i \partial_i U \right\rangle, 
 \label{dQ-ex-1}
\ea
where $\langle \, \cdot \, \rangle$ means average over the pdf $p(\xv, t)$  of $\xv$:
\ba
\langle \, \cdot \, \rangle 
 = \int_\xv \cdot \,  p(\xv, t) . 
\ea
Similarly, the differential housekeeping heat at the ensemble level is given by 
\ba
\dbar Q^{\rm hk} =  - \frac{T^2\, dt}{\gamma} 
\left\langle 
\varphi_i (\partial_i U - \varphi_i) 
- \partial_i \varphi_i
 \right\rangle.
\ea
Further using the Gibbs gauge condition (\ref{Gibbs-gauge-cond}) we may rewrite the above result as
\ba
\dbar Q^{\rm hk} =  - \frac{T^2\, dt}{\gamma} 
 \int_\xv p (\varphi_i)^2 \leq 0,
\ea
which is non-positive definite. 

The differential housekeeping and excess work at the trajectory level can be similarly computed:
\ba
\dbar W^{\rm hk} &=& -  \dbar Q^{\rm hk} 
=  \frac{T^2\, dt}{\gamma} 
 \int_\xv p (\varphi_i)^2 \geq 0, \\
 \dbar W^{\rm hk} &=& T \int_\xv  p \, d_\lambda U
\ea

The total heat and work at the ensemble level are then the sum of the corresponding housekeeping parts and excess parts:
\ba
\dbar { Q} &=& 
 \langle \!\langle \dbar {\mathscr Q} \rangle \! \rangle
 = \dbar { Q}^{\rm hk} + \dbar { Q}^{\rm ex},\\
\dbar { W} &=&
 \langle \!\langle \dbar {\mathscr W} \rangle \! \rangle
=  \dbar { W}^{\rm hk} + \dbar { W}^{\rm ex}. 
\ea

The system entropy is:
\ba
S[p] = - \int_\xv p \, \log p, 
\ea
whose differential can be calculated using the Fokker-Planck equation:
\ba
dS &=& - dt \int_\xv \log p\, \mathscr L p
\nonumber\\
&=&  - dt \int_\xv \log p\,  \partial_i \frac{T}{\gamma} 
    (\partial_i + \partial_i U  - \varphi_i ) p
\nonumber\\
&=& \frac{dt\, T}{\gamma} \int_\xv \frac{1}{p} (\partial_i p ) 
(\partial_i  + \partial_i U - \varphi_i) p,
\label{dS-1-1}
\ea
where in the last step we have integrated by parts. 

The  EP  is defined as 
\ba
  d S^{\rm tot} &\equiv& dS + dS^{\rm env}  
=  dS - \beta \, \dbar Q,
\ea
where $dS^{\rm env}   \equiv - \beta \, \dbar Q$ is {\em defined} as the environmental entropy change.  {As explained previously,  $dS^{\rm env}$ is only the part of environmental entropy change that can be captured by our theory of stochastic thermodynamics. }
This can be further decomposed into a housekeeping  EP  and an excess  EP :
\ba
  d S^{\rm tot} &= & dS^{\rm hk} +  dS^{\rm ex},\\
  dS^{\rm hk} &=& - \beta \dbar Q^{\rm hk} 
  =   \frac{T\, dt}{\gamma}   \int_\xv p (\varphi_i)^2   \geq 0, \\
   dS^{\rm ex} &=& dS  - \beta \, \dbar Q^{\rm ex} .
\ea
In particular, in the NESS, the excess  EP  vanishes identically, whereas the housekeeping  EP  reduces to 
\ba
\frac{dS^{\rm hk}}{dt} =  \frac{T\, dt}{\gamma}   \int_\xv e^{- U} (\varphi_i)^2 
= \frac{\gamma}{T}   \int_\xv e^U \left( j_i^{\rm ss} \right)^2,
\ea
where $j_i^{\rm ss}$ is the NESS current given in Eq.~(\ref{j^ss-def-1}). 

Further using Eqs.~(\ref{dS-1-1}) and (\ref{dQ-ex-1}), as well as the Gibbs gauge condition (\ref{Gibbs-gauge-cond}), we may rewrite the excess  EP  in the following apparently positive form:
\ba
dS^{\rm ex}  = \frac{T\, dt}{\gamma}   \int_\xv
\frac{1}{p} \left[ (\partial_i  + \partial_i U) p\right]^2
\geq 0.
\ea
Hence  EP  is the sum of a positive housekeeping part and a positive excess part, a general feature of Markov systems with even variables and parameters that lack instantaneous detailed balance~\cite{chernyak_2006_pathintegral,esposito_2010_three,esposito_2010_threea,vandenbroeck_2010_three}. 

Finally we may also define non-equilibrium free energy:
\ba
F[p] \equiv \int_\xv p ( H + T \, \log p).
\ea
It is then easy to verify the following differential forms:
\ba
d  F[p] = \dbar W^{\rm ex} + \dbar Q^{\rm ex} - T \,dS,
\ea
which may be further rewritten as 
\ba
dS^{\rm ex} =  dS  - \beta \, \dbar Q^{\rm ex}
= \beta ( \dbar W^{\rm ex}  -  d  F[p])
\geq 0. 
\ea

\subsection{Transition probability}\label{appendix::probability}
\vspace{-3mm}
To study fluctuation theorems, it is necessary to know the short-time transition probability of  the Langevin process defined by Eq.~(\ref{Langevin-Gibbs}).  Let  $\xv, \xv_1 = \xv + d \xv$ be respectively the initial position and the final position of an infinitesimal transition taking place during $dt$, and let $\xv_{1/2} = \xv + d \xv/2$ be the mid-point.  A general expression for the short-time transition probability $p_{\bm \varphi}(\xv_1 |\xv; dt) $ of the Langevin equation (\ref{Langevin-4}) was derived in Eqs.~(A4) of Ref.~\cite{ding_unified_2022}, using the general result of time-slicing path integral in Ref.~\cite{ding_time-slicing_2022}.  Specializing to the Langevin dynamics Eq.~(\ref{Langevin-Gibbs}), we find~\footnote{The notation may be unfortunately confusing since we need to integrate over $d\xv$ to verify the normalization of this transition probability density function.  To avoid this confusion, we may replace $d\xv, dt$ by $\Delta \xv, \Delta t$ and remember that they are infinitesimal quantities. }  (Note that the notations are slightly different here)
\begin{subequations}
\ba
p_{\bm \varphi}(\xv_1 |\xv; dt) = \frac{\gamma}{4\pi T\, dt} 
e^{- {\mathcal A}_{\bm \varphi}(d \xv; \xv_{1/2}, dt)},
  \label{dx-pdf-2}
\ea 
where {\em the action} ${\mathcal A}_{\bm \varphi}(d \xv; \xv_{1/2}, dt)$ is given by 
\ba
{\mathcal A}_{\bm \varphi}(d \xv; \xv_{1/2}, dt) &=& \frac{\gamma}{4 T\, dt} 
  (dx^i + \frac{T}{\gamma} (\partial_i U - \varphi_i) dt)^2_{1/2}
  \nonumber\\
 &-& \frac{T\, dt}{2 \gamma} 
 (\partial_i^2 U - \partial_i \varphi_i)_{1/2}
 + o(dt), 
 \nonumber\\
 \label{action-1/2} \quad\quad
\ea
\label{transition-pdf-action-1/2}
\end{subequations}
where the subscript $1/2$ in Eq.~(\ref{action-1/2}) means that all functions inside the bracket are evaluated at $\xv_{1/2}$.   The action is expanded up to the first order in $dt$, which is sufficient to guarantees a correct continuum limit. In fact it is ok to evaluate the second term in the r.h.s. of Eq.~(\ref{action-1/2}) at any point, the resulting error is of higher order than $dt$, and hence is negligible in the continuum limit. Note that we show explicitly the dependence of the action on the non-conservative force $\bm \varphi$.

Let us supply a heuristic explanation for Eqs.~(\ref{transition-pdf-action-1/2}).  First we note that the Wiener noises are infinitesimal Gaussian with basic properties (\ref{Wiener-properties}).  Using these we can readily construct their pdf:
\begin{align}
    p(d\bm W) = \frac{1}{{2\pi dt}}\exp \left(
    -\frac{dW_x^2  + dW_y^2}{2 dt} \right). 
    \label{p-dW}
\end{align}
Now given $\xv$, the Langevin equation (\ref{Langevin-Gibbs}) may be understood as a linear relation between $d \bm W$ and the infinitesimal displacement $d\xv$.  Hence we may obtain the pdf for $d\xv$ directly from Eq.~(\ref{p-dW}):
\ba
p(d\xv)  = \frac{\gamma}{4\pi T\, dt} 
\, e^ { - \frac{\gamma}{4\pi T\, dt} 
  (dx^i + \frac{T}{\gamma} (\partial_i U - \varphi_i) dt)^2 }.
  \label{dx-pdf-1}
\ea
Note that the action appearing in the exponent above is formally identical to the first term in the action (\ref{action-1/2}).   It is important to note however as a basic property of Ito-Langevin equation,  the function  $  ( \partial_i U - \varphi_i ) $ in  (\ref{Langevin-Gibbs}),  which also appears in Eq.~(\ref{dx-pdf-1}) is evaluated at the initial point $\xv$.   This should be contrasted with Eqs.~(\ref{transition-pdf-action-1/2}), where the same function is evaluated at the mid-point $\xv_{1/2}$.  Because of the $dt$ appearing in the denominator of the actions, however, this difference is qualitatively important and is compensated by the second term in the action (\ref{action-1/2}).  

Using Eq.~(\ref{dx-pdf-2}), we can also compute the backward transition probability from $\xv_1$ to $\xv$.  All we need is to swap $\xv_1$ and $\xv$ in Eq.~(\ref{dx-pdf-2}).  Note that $\xv$ and $\xv_{1}$ appear in  Eq.~(\ref{dx-pdf-2}) only in the combinations $d\xv$ and $\xv_{1/2}$, which are respectively odd and even under the swap.   Hence to obtain $p(\xv |\xv_1; dt)$ we only need to flip the sign of $d\xv$.  This leads to 
\begin{subequations}
\ba
p_{\bm \varphi}(\xv |\xv_1; dt) &=& \frac{\gamma}{4\pi T\, dt} 
e^{- {\mathcal A}_{\bm \varphi}( - d \xv; \xv_{1/2}, dt)},
  \label{dx-pdf-4}\\
{\mathcal A}_{\bm \varphi}( - d \xv; \xv_{1/2}, dt) &=& \frac{\gamma}{4 T\, dt} 
  ( - dx^i + \frac{T}{\gamma} (\partial_i U - \varphi_i) dt)^2_{1/2}
  \nonumber\\
 &-& \frac{T\, dt}{2 \gamma} 
 (\partial_i^2 U - \partial_i \varphi_i)_{1/2}
 + o(dt). \nonumber\\
 \label{action-1/2-flip} \quad\quad
\ea
\label{transition-pdf-action-1/2-flip}
\end{subequations}
Recall the adjoint process defined in Sec.~\ref{sec:adjoint} is related to the original process by changing the sign of $\bm \varphi$.  We can construct the corresponding transition probability for the adjoint process from Eqs.~(\ref{transition-pdf-action-1/2}):
\begin{subequations}
\ba
p_{- \bm \varphi}(\xv_1 |\xv; dt) &=& \frac{\gamma}{4\pi T\, dt} 
e^{- {\mathcal A}_{- \bm \varphi}(d \xv; \xv_{1/2}, dt)},
  \label{dx-pdf-2-Ad}\\
{\mathcal A}_{- \bm \varphi}(d \xv; \xv_{1/2}, dt) &=& \frac{\gamma}{4 T\, dt} 
  (dx^i + \frac{T}{\gamma} (\partial_i U +  \varphi_i) dt)^2_{1/2}
  \nonumber\\
 &-& \frac{T\, dt}{2 \gamma} 
 (\partial_i^2 U + \partial_i \varphi_i)_{1/2}
 + o(dt), 
 \nonumber\\
 \label{action-1/2-Ad} \quad\quad
\ea
\label{transition-pdf-action-1/2-Ad}
\end{subequations}
The backward transition probability of the adjoint process can be similarly obtained from Eqs.~(\ref{transition-pdf-action-1/2-flip}):
\begin{subequations}
\label{transition-pdf-action-1/2-flip-Ad}
\ba
p_{-\bm \varphi}(\xv |\xv_1; dt) &=& \frac{\gamma}{4\pi T\, dt} 
e^{- {\mathcal A}_{- \bm \varphi}( - d \xv; \xv_{1/2}, dt)},
  \label{dx-pdf-4-Ad}\\
{\mathcal A}_{- \bm \varphi}( - d \xv; \xv_{1/2}, dt) &=& \frac{\gamma}{4 T\, dt} 
  ( - dx^i + \frac{T}{\gamma} (\partial_i U + \varphi_i) dt)^2_{1/2}
  \nonumber\\
 &-& \frac{T\, dt}{2 \gamma} 
 (\partial_i^2 U + \partial_i \varphi_i)_{1/2}
 + o(dt). \nonumber\\
 \label{action-1/2-flip-Ad} \quad\quad
\ea
\end{subequations}

Using Eqs.~(\ref{transition-pdf-action-1/2}) and Eqs.~(\ref{transition-pdf-action-1/2-flip})-(\ref{transition-pdf-action-1/2-flip-Ad}), we readily obtain the following ratios:
\begin{subequations}
\ba
\frac{p_{\bm \varphi}(\xv_1 |\xv; dt)}{p_{ \bm \varphi}(\xv |\xv_1; dt) } 
&=& e^{- (\partial_i U - \varphi_i) \circ dx^i} .
\label{cond-generalized-DB-1}
\ea
 Similarly, with the aid of Gibbs gauge condition Eq.~(\ref{Gibbs-gauge-cond}), we may also prove 
\ba
\frac{p_{\bm \varphi}(\xv_1 |\xv; dt)}{p_{- \bm \varphi}(\xv_1 |\xv; dt) } 
&=& \frac{p_{-\bm \varphi}(\xv |\xv_1; dt)}{p_{  \bm \varphi}(\xv |\xv_1; dt) } 
=  e^{\bm \varphi \circ d \xv},
\label{cond-generalized-DB-2}
 \\
\frac{p_{\bm \varphi}(\xv_1 |\xv; dt)}{p_{- \bm \varphi}(\xv |\xv_1; dt) } 
&=& \frac{p_{-  \bm \varphi}(\xv_1 |\xv; dt)}{p_{ \bm \varphi}(\xv |\xv_1; dt) } 
=  e^{ - d_\xv U}.  \quad \quad
\label{cond-generalized-DB-3}
\ea
\label{cond-generalized-DB}
\end{subequations}
Equations (\ref{cond-generalized-DB}) may be called {\em the conditions of local detailed balance}. 

\subsection{Four processes}
\label{sec:4-processes}

\begin{table*}[t!]
    \centering
    \setlength{\tabcolsep}{3pt}
    \renewcommand{\arraystretch}{1.25}
    \begin{tabular}{|c|c|c|c|c|}
        \hline
        Process: & \begin{tabular}[c]{c} Generalized \\ potential \end{tabular}
         & \begin{tabular}[c]{c} Non-conservative \\ force \end{tabular} 
         & Confining potential & Velocity field         \\ \hline
        Forward & $U(\xv, \lambda_t )$ 
        & $\bm \varphi(\xv, \lambda_t)$  
        & $V(\xv; \lambda_t)$
        & $\bm v(\xv)$
        \\ \hline
        Backward & $U(\xv, \lambda_{\tau - t} )$ 
        & $\bm \varphi(\xv, \lambda_{\tau - t})$
        & $V(\xv; \lambda_{\tau - t})$
        & $\bm v(\xv)$ 
          \\ \hline
        Adjoint Forward
         & $U(\xv, \lambda_t )$ 
         & $ - \bm \varphi(\xv, \lambda_t)$  
        & $V(\xv; \lambda_t) + 2 \, T\, \psi(\xv; \lambda_t)$
        & $ -\bm v(\xv)$
                 \\ \hline
        Adjoint Backward
        & $U(\xv, \lambda_{\tau - t} )$ 
        & $ - \bm \varphi(\xv, \lambda_{\tau - t})$  
        & $V(\xv; \lambda_{\tau - t})       
         + 2 \, T\, \psi(\xv; \lambda_{\tau - t})$
        &$ - \bm v(\xv)$
        \\ \hline
    \end{tabular}
    \caption{ Dynamic protocols of all four processes.  Column 2 and 3 show the protocols in the Gibbs gauge, whereas Column 4 and 5 show the protocols in the natural gauge.   }
    \label{table::Process for fluctuation theorem}
\vspace{-3mm}
\end{table*}

We keep the flow field fixed, and vary  parameters $\lambda_t = \{ K(t), \xv_0(t) \}$ systematically, which fully determines the Langevin dynamics.  We call $\{U(\xv; \lambda_t), \bm \varphi (\xv; \lambda_t) \}$ {\em the dynamic protocol in the Gibbs gauge}, and $\{V(\xv; \lambda_t), \bm v(\xv)\}$ {\em the dynamic protocol in the natural gauge}.  
A dynamic process is determined by the initial pdf of $p(\xv, t=0)$ together with the dynamic protocol either in the Gibbs gauge or in the natural gauge.  Transformation between two dynamic protocols are given by Eqs.~(\ref{gauge-transform}) and (\ref{U-varphi-def}).

We define four processes as below, all of which start from $t = 0$ and end at $t = \tau$:
\begin{enumerate}
\item {\em Forward process:} \,  The dynamic protocol is 
\begin{subequations}
 \label{equ::Forward process}
\ba
U^{\rm F}&=& U(\xv, \lambda_t), \quad \\
\bm \varphi^{\rm F} &=& \bm \varphi(\xv, \lambda_t).
\ea
\end{subequations}
The initial pdf is $p^{\rm ss}(\xv; \lambda_0)$, defined in Eq.~(\ref{p^ss-1}). 

\item {\em Backward process:} \, The dynamic protocol is 
\begin{subequations}
\label{equ::Backward process}
\ba
U^{\rm B} &=&  U(\xv, \lambda_{\tau - t}), \quad \\
\bm \varphi^{\rm B} &=&  \bm \varphi(\xv, \lambda_{\tau - t}). 
\ea
\end{subequations}
 The initial pdf is $p^{\rm ss}(\xv; \lambda_\tau)$.   
 
\item {\em Adjoint process:} \, The dynamic protocol is 
\begin{subequations}
\label{equ::Adjoint process}
\ba
U^{\rm Ad} &=&  U(\xv, \lambda_t), \quad \\
\bm \varphi^{\rm Ad} &=& - \bm \varphi(\xv, \lambda_t). 
\ea
\end{subequations}
The initial pdf is $p^{\rm ss}(\xv; \lambda_0)$.  

\item {\em Adjoint backward process:}  \, The  protocol is 
\begin{subequations}
 \label{equ::Adjoint backward process}
\ba
U^{\rm AdB} &=&  U(\xv, \lambda_{\tau - t}), \quad \\
\bm \varphi^{\rm AdB} &=& - \bm \varphi(\xv, \lambda_{\tau - t}).
\ea
\end{subequations}
The initial pdf is $p^{\rm ss}(\xv; \lambda_\tau)$. 
\end{enumerate}
Note that each of these processes starts from the NESS corresponding to the initial control parameter of the dynamic protocol. Such an initial state can be realized easily in experiments. Note also that in general, the system is not in a NESS at the end of any of these processes. 

The protocols of all these processes are displayed in the second and third columns of Table \ref{table::Process for fluctuation theorem}. We may also express these protocols in the natural gauge, in terms of the confining potential and the flow field.  The results are displayed in fourth and fifth columns of Table \ref{table::Process for fluctuation theorem}.

A pivotal property of these processes is that the backward process, the adjoint process, and the adjoint backward process are all uniquely determined by the forward process.  Furthermore, the backward of the backward process is the forward process.  Likewise, the adjoint of the adjoint process is the forward process; the adjust backward of the adjoint backward process is also the forward process.  Additionally, the adjoint of the backward process is the same as the backward of the adjoint process, which is also the same as the adjoint backward process etc. The mappings from any process to its backward process, and that to its adjoint process, as well as that to its adjoint backward process, are all involutions. 

Consider a  trajectory and its backward trajectory:
\ba
\bm \gamma &=& \{\xv(t), t \in [0, \tau] \},\\
\hat {\bm \gamma} &=&  \{\xv(\tau - t), t \in [0, \tau] \}. 
\ea 
The notation $\bm \gamma$ (boldface) for trajectory should be carefully distinguished from $\gamma$ for the friction coefficient.  We introduce the notations ${\bm \gamma}_0  = \xv(0)$ and $\hat {\bm \gamma}_0 = \xv(\tau)$ to denote the initial state of $\bm \gamma, \hat {\bm \gamma}$, respectively.  For each of the four processes defined above, we can construct its pdf of trajectory as the product of conditional pdf given the initial state and the pdf of the initial state.  For example, for the forward process, we have
\begin{subequations}
\label{p-gamma}
\ba
p_{\rm F}[ \bm \gamma] = p_{\rm F}[\bm \gamma | {\bm \gamma}_0] \,
 p^{\rm ss}(\xv(0); \lambda_0). 
 \label{p_F-gamma}
\ea
   Similarly, we also have for the other three processes:
\ba
p_{\rm B}[\hat {\bm \gamma}] &=& 
p_{\rm B}[\hat {\bm \gamma} |\hat  {\bm \gamma}_0]  \,
p^{\rm ss}(\xv(\tau); \lambda_\tau),
 \label{p_B-gamma}\\
p_{\rm Ad}[ \bm \gamma] &=& 
p_{\rm Ad}[\bm \gamma | {\bm \gamma}_0] \,
p^{\rm ss}(\xv(0); \lambda_0) ,
 \label{p_Ad-gamma}\\
p_{\rm AdB}[\hat {\bm \gamma}] &=& 
p_{\rm AdB}[\hat {\bm \gamma} |\hat  {\bm \gamma}_0] \,
p^{\rm ss}(\xv(\tau); \lambda_\tau).  
 \label{p_AdB-gamma}
\ea
\end{subequations}

Let  $\mathscr W_{\rm F} [ \bm \gamma], \mathscr Q_{\rm F} [ \bm \gamma]$ ($\mathscr W_{\rm B} [\hat {\bm \gamma}],\mathscr Q_{\rm B} [\hat {\bm \gamma}] $) be the integrated work and heat along $\bm\gamma$ ($\hat {\bm \gamma}$) in the forward (backward) process.  
They can be obtained by integrating Eqs.~(\ref{equ::W-def}) and (\ref{dQ-traj-1}) along the forward (backward) trajetory:
\begin{subequations}
\ba
\mathscr W_{\rm F} [ \bm \gamma] = 
- \mathscr W_{\rm B} [\hat {\bm \gamma}] 
= T  \int_\gamma   ( d_\lambda U  + \bm \varphi \circ d \xv ),
\label{W-F-decomp-1}\\
\mathscr Q_{\rm F} [ \bm \gamma] = 
- \mathscr Q_{\rm B} [\hat {\bm \gamma}] 
=  T  \int_\gamma  ( d_{\bm x} U  -  \bm \varphi \circ d \xv ) . 
\label{Q-F-decomp-1}
\ea
We can similarly construct the same quantities for the adjoint process and the adjoint backward process:
\ba
\mathscr W_{\rm Ad} [ \bm \gamma] = 
- \mathscr W_{\rm AdB} [\hat {\bm \gamma}] 
&=& T  \int_\gamma   ( d_\lambda U  - \bm \varphi \circ d \xv ),
\label{W-Ad-decomp-1}\\
\mathscr Q_{\rm Ad} [ \bm \gamma] = 
- \mathscr Q_{\rm AdB} [\hat {\bm \gamma}] 
&=&  T  \int_\gamma  ( d_{\bm x} U  + \bm \varphi \circ d \xv ) . 
\label{Q-Ad-decomp-1} \quad\quad\quad
\ea
\end{subequations}

The integrated work and heat may be decomposed into a housekeeping part and an excess part. For the work of the forward process, we have:
\begin{subequations}
\ba
\mathscr W_{\rm F} [ \bm \gamma] &=& 
\mathscr W_{\rm F}^{\rm hk} [ \bm \gamma] 
+ \mathscr W_{\rm F}^{\rm ex} [ \bm \gamma],
\label{W-hk-ex-def}\\
\mathscr W_{\rm F}^{\rm hk} [ \bm \gamma]  &=& 
T \int_{\bm \gamma}  \bm \varphi \circ d \xv,\\
\mathscr W_{\rm F}^{\rm ex} [ \bm \gamma]   &=& 
T \int_{\bm \gamma} d_\lambda U.
\ea
\end{subequations}
The heat of the forward process can be decomposed in a similar way. 
Same decompositions can also be obtained for work and heat of the backward process, the adjoint process, and the adjoint backward process.  

Combining, we obtain
\ba
\mathscr W^{\rm hk}_{\rm F}[ \bm \gamma] &=&
 - \mathscr W^{\rm hk}_{\rm B}[\hat{\bm \gamma}] 
= - \mathscr W^{\rm hk}_{\rm Ad}[ \bm \gamma]
=  \mathscr W^{\rm hk}_{\rm AdB}[\hat{\bm \gamma}]
\nonumber\\
&=& T \int_{\bm \gamma}  \bm \varphi \circ d \xv,
\label{w-hk-total-sym}
\\
\mathscr W^{\rm ex}_{\rm F}[ \bm \gamma] &=&
 - \mathscr W^{\rm ex}_{\rm B}[\hat{\bm \gamma}] 
=  \mathscr W^{\rm ex}_{\rm Ad}[ \bm \gamma]
=  - \mathscr W^{\rm ex}_{\rm AdB}[\hat{\bm \gamma}]
\nonumber\\
 &=&  T \int_{\bm \gamma} d_\lambda U.
\label{w-ex-total-sym}
\ea

Finally, adding up Eqs.~(\ref{W-F-decomp-1}) and (\ref{Q-F-decomp-1}),  we obtain the first law along $\bm \gamma$:
\ba
U(\xv(\tau); \lambda_\tau) - U(\xv(0) ; \lambda_0)
= \mathscr W_{\rm F} [ \bm \gamma] 
+ \mathscr Q_{\rm F} [ \bm \gamma]. 
\label{1st-law-int}
\ea

\subsection{Fluctuation theorems}

Because of the Markov property, $p_{\rm F}[\bm \gamma  |{\bm\gamma}_0]$ and $ p_{\rm B}[\hat {\bm \gamma} |\hat  {\bm\gamma}_0] $ can be calculated using the time-slicing method.   Further using Eq.~(\ref{cond-generalized-DB-1}) for each pair of time-slices, we have
\ba
\log  \frac{ p_{\rm F}[\bm \gamma | {\bm\gamma}_0] }  
{ p_{\rm B}[\hat {\bm \gamma} |\hat  {\bm\gamma}_0] } 
= - \int_\gamma  ( d_{\bm x} U  -  \bm \varphi \circ d \xv )
= {- \beta \mathscr Q_{\rm F}[ \bm \gamma]},
\label{p-gamma-cond-ratio} \quad
\ea
where in the second equality we have used Eq.~(\ref{Q-F-decomp-1}).


Let us define the following functional: 
\ba
\Sigma_{\rm F}[ \bm \gamma] \equiv
  \log \frac{p_{\rm F}[ \bm \gamma]}{p_{\rm B}[\hat {\bm \gamma}]} .
\label{p-gamma-ratio-0}
\ea
Using Eqs.~(\ref{p_F-gamma}), (\ref{p_B-gamma}), and (\ref{p-gamma-cond-ratio}), we obtain:
  \ba
\Sigma_{\rm F}[ \bm \gamma] 
=  \log \frac{ p^{\rm eq}(\xv(0); \lambda_0)}
{p^{\rm eq}( \xv(\tau); \lambda_\tau)}
-  \beta \mathscr Q_{\rm F}[ \bm \gamma].
\quad \label{p-gamma-ratio-1} 
\ea
If the final state pdf $p(\xv(\tau), \tau)$ of the forward process is the NESS corresponding to $\lambda_\tau$, we may also write Eq.~(\ref{p-gamma-ratio-1}) into 
\ba
\Sigma_{\rm F}[ \bm \gamma] = 
- \log \frac{ p(\xv(\tau), \tau)}{p(\xv(0), 0)}
- \beta \mathscr Q_{\rm F} [ \bm \gamma],
\label{Sigma-gamma-1}
\ea
which is {\em the stochastic entropy production} along the trajectory $\bm \gamma$ in the forward process.  If the system is not in the NESS at the end of the forward process, however, the physical meaning of $\Sigma_{\rm F}[ \bm \gamma] $ is more subtle. 

Further taking advantage of Eq.~(\ref{p^ss-1}) as well as  Eqs.~(\ref{1st-law-int}) and (\ref{W-F-decomp-1}), we may rewrite Eq.~(\ref{p-gamma-ratio-1}) into:
\ba
\Sigma_{\rm F}[ \bm \gamma] = 
\log \frac{p_{\rm F}[ \bm \gamma]}{p_{\rm B}[\hat {\bm \gamma}]} 
=\beta \mathscr W_{\rm F}[\bm \gamma]
=  - \beta \mathscr W_{\rm B} [\hat {\bm \gamma}].
\label{p-gamma-ratio-1-1}
\ea

Taking the log ratio of Eqs.~(\ref{p_F-gamma}) and (\ref{p_Ad-gamma}) we find
\ba
\log \frac{p_{\rm F}[ \bm \gamma]}{p_{\rm Ad}[ {\bm \gamma}]} 
=\log \frac{p_{\rm F}[\bm \gamma | {\bm \gamma}_0] }
{p_{\rm Ad}[\bm \gamma | {\bm \gamma}_0] }.
\ea
The r.h.s. can be calculated using the time-slicing method and Eq.~(\ref{cond-generalized-DB-2}).  The result is 
\ba
\log \frac{p_{\rm F}[ \bm \gamma]}{p_{\rm Ad}[ {\bm \gamma}]} 
= \int_{\bm \gamma} \bm \varphi \circ d \xv. 
\ea
Similarly we may also obtain:
\ba
 \log  \frac{p_{\rm AdB}[\hat {\bm \gamma}]}
 {p_{\rm B}[\hat{\bm \gamma}]} 
= \log  \frac{p_{\rm AdB}[\hat {\bm \gamma}|\hat{\bm \gamma_0}]}
{p_{\rm B}[\hat{\bm \gamma}|\hat{\bm \gamma_0}]} 
= \int_{\bm \gamma} \bm \varphi \circ d \xv.
\ea
Combining the preceding two equations, and further using Eq.~(\ref{w-hk-total-sym}), we obtain 
\ba
\log \frac{p_{\rm F}[ \bm \gamma]}{p_{\rm Ad}[ {\bm \gamma}]} 
&=& \log  \frac{p_{\rm AdB}[\hat {\bm \gamma}]}{p_{\rm B}[\hat{\bm \gamma}]} 
= \beta \mathscr W^{\rm hk}_{\rm F} [ \bm \gamma]
= -  \beta \mathscr W^{\rm hk}_{\rm Ad} [ \bm \gamma];
\nonumber\\
\label{DFT-W-W-hk}
\ea
\begin{subequations}
Finally using similar methods, we may also prove
\ba
\log \frac{p_{\rm F}[ \bm \gamma]}{p_{\rm AdB}[ {\hat {\bm \gamma}}]} 
&=& \log  \frac{p_{\rm Ad}[ {\bm \gamma}]}{p_{\rm B}[\hat{\bm \gamma}]} 
= \beta \mathscr W^{\rm ex}_{\rm F} [ \bm \gamma]
= -  \beta \mathscr W^{\rm ex}_{\rm AdB} [ \bm \gamma]. 
\nonumber\\
\label{DFT-W-W_ex}
\ea
\end{subequations}

Let us now define the pdf of $ \mathscr W_{\rm F}[ \bm \gamma], \mathscr W_{\rm F}^{\rm hk} [ \bm \gamma], \mathscr W_{\rm F}^{\rm ex} [ \bm \gamma]$ for the forward process  as:
\begin{subequations}
\ba
p_{\rm F}(  \mathscr W) &\equiv& \int D {\bm \gamma} \, 
\delta \left(  \mathscr W -  \mathscr W_{\rm F}[ \bm \gamma] \right) \, p_{\rm F}[ \bm \gamma], \\
p_{\rm F}(  \mathscr W^{\rm hk}) &\equiv& \int D {\bm \gamma} \, 
\delta \left(  \mathscr W^{\rm hk} -  \mathscr W_{\rm F}^{\rm hk}
[ \bm \gamma] \right) \, p_{\rm F}[ \bm \gamma], \quad\quad\\
p_{\rm F}(  \mathscr W^{\rm ex} ) &\equiv& \int D {\bm \gamma} \, 
\delta \left(  \mathscr W^{\rm ex}  -  \mathscr W_{\rm F}^{\rm ex}
 [ \bm \gamma] \right) \, p_{\rm F}[ \bm \gamma],
\ea
\end{subequations}
where $\int D  \gamma$ means functional integration in the space of dynamic trajectories. This functional integral should be computed using time-slicing, similar to the path integral in quantum mechanics.   Similar pdfs can also be defined for the work, the housekeeping work, and the excess work in the backward process, the adjoin process, and the adjoint backward process.

Taking advantage of the symmetries (\ref{p-gamma-ratio-1-1}), (\ref{DFT-W-W-hk}), and (\ref{DFT-W-W_ex}), and using standard methods of stochastic thermodynamics, we can prove the following {\em fluctuation theorems} for the work, the housekeeping work, and the excess work:
\begin{subequations}
\label{FT-all}
\ba
p_{\rm F}(\mathscr W)  &=& e^{\beta \mathscr W} p_{\rm B}(- \mathscr W), \label{equ::FT of tot work} \\
p_{\rm F}(\mathscr W^{\rm hk}) &=& e^{\beta \mathscr W^{\rm hk}} p_{\rm Ad}(- \mathscr W^{\rm hk}), \label{equ::FT of hk work} \\
p_{\rm F}( \mathscr W^{\rm ex})  &=& e^{\beta \mathscr W^{\rm ex}} p_{\rm AdB}(- \mathscr W^{\rm ex}).  \label{equ::FT of ex work}
\ea
\end{subequations}

\section{Alternative theory}
\label{sec:difference}

Here we briefly review the theory of Speck {\it e. al.}~\cite{speck_2008_role}, which was established on the same Langevin dynamics  (\ref{langevin-1}).  We shall compare two theories and highlight their differences. 

Noticing that the concepts of heat in stochastic thermodynamics is not Galilean invariant, the authors of Ref.~\cite{speck_2008_role} argue that one should transform to the co-moving frame and implement the usual formalism of stochastic thermodynamics.  For obvious reasons, let us call this theory {\em the theory of co-moving frames}.
The heat is therefore defined as negative the work done by the friction and random forces in the co-moving frame.  Using the Langevin equation (\ref{langevin-1}), we find:
\begin{align}
    \dbar \mathscr{Q}^{\rm cm}  &\equiv 
     -\left[ \gamma\left( \frac{d \xv}{dt }- {\bm v} \right) 
     - \sqrt{2 \gamma T} d \bm W\right]
      \circ \left(d \xv - {\bm v} d t\right)
    \nonumber \\
        & =  \bm \nabla V \circ  \left(d \xv - {\bm v} d t\right) 
         \label{shear-flow-dq}\\
    & \equiv  d_{\boldsymbol{x}} V - \bm v \circ \bm \nabla V d t,
\nonumber   
\end{align}
where the superscript { cm} denotes co-moving.  Note however, for a shear flow, the co-moving frame is not an inertial frame. 

The heat at the ensemble level can be computed using the same method as we used in Sec.~\ref{sec:W-Q-Sigma}. The result is 
\begin{align}
\dbar Q^{\rm cm} &= \langle\! \langle \dbar \mathscr Q^{\rm cm} \rangle \! \rangle
\nonumber \\
&=  -dt \int_{\xv} (\partial_i V)  \frac{T}{\gamma}
 (\partial_i + \beta \partial_i V )p. 
\end{align}
The   EP  in the co-moving theory is:
\begin{align}
 dS^{\rm cm, tot} & = dS^{\rm sys} - \beta \dbar Q^{\rm cm}
 \label{dS-tot-cm}
\\
&= \frac{T dt }{\gamma}
\int_{\xv} \frac{1}{p }( \partial_i p +\beta p \,\partial_i V )^2 
+ dt \int_\xv (\bm \nabla \cdot \bm v) p,
\nonumber
\end{align}
where we have used Eq.(\ref{dS-1-1}).  Whereas the first term in the r.h.s. of Eq.~(\ref{dS-tot-cm}) is non-negative, the second term does not have a definite sign, and vanishes only if the fluid is incompressible.  Hence if the fluid is compressible, the EP in the co-moving theory is not necessarily positive. 

Assuming that the fluid is incompressible, Eq.~(\ref{dS-tot-cm}) becomes 
\begin{align}
 dS^{\rm cm, tot} &= \frac{T dt }{\gamma}
\int_{\xv} \frac{1}{p }( \partial_i p +\beta p \,\partial_i V )^2 
\geq 0.
 \label{dS-tot-cm-1}
\end{align}
Note that this  EP  vanishes identically if the pdf is Gibbs-Boltzmann with respect to the external potential: $p \sim e^{-\beta V}$.  Such a state, however, is not the NESS of the Langevin dynamics.

The fluctuating internal energy is defined as the external potential $V$.  By imposing the first law of thermodynamics:
\ba
d V = \dbar \mathscr{W}^{\rm cm}   + \dbar \mathscr{Q}^{\rm cm} , 
\ea
one finds that the work at the trajectory level is 
\begin{align}
\dbar \mathscr{W}^{\rm cm}  & =d V - \dbar \mathscr{Q}^{\rm cm} 
\nonumber \\
& =  d_{\lambda} V + \bm v \circ \bm \nabla V d t,
\label{equ::d work cm}
\end{align}

The conditions of local detailed balance (\ref{cond-generalized-DB}), which relate the transition probabilities of the forward and backward processes to the heat exchange between the system and the environment, play an essential role in the theory of stochastic thermodynamics.  It turns out that the heat defined by Eq.~(\ref{shear-flow-dq}) is also related to a similar condition concerning a different definition of backward process.   This backward process is characterized by the reversal of both the time-variable and the flow field.  In other words, the backward process in the co-moving theory is defined such that the dynamic protocol is $\lambda_{\tau - t}$, whereas the flow field is $- \bm v(\xv; \lambda_{\tau - t})$.   The probability of the backward transition in the backward process, denoted using the superscript $*$, is then
\begin{align}
p^*(\xv_0|\xv_1,dt) & = e^{- A^*(\xv_1|\xv_0,dt) } , \\
 A^*(\xv_0|\xv_1,dt) &= \frac{1}{4 T \gamma dt} ( - \gamma dx^i + \gamma v_i dt +  \partial_i V dt)^2 
\nonumber\\
  &- \frac{1}{2\gamma} (\partial_i^2 V - \partial_i \gamma v_i) \, dt.
\label{reverse-action}
\end{align}
If we take the ratio of the transition probabilities of the forward and backward processes, we obtain
\ba
\log\frac{p(\xv_1|\xv_0,dt) }{ p^*(\xv_0|\xv_1,dt)} 
 = - \beta \dbar \mathscr Q^{\rm cm},
\label{dft-def1}
\ea
where $ \dbar \mathscr Q^{\rm cm}$ is defined by Eq.~(\ref{shear-flow-dq}).  This is the condition of local detailed balance for the theory of co-moving frame. 

If we choose the initial states of the forward process and the backward process to be equilibrium states (with the flow field completely turned off):
\begin{subequations}
\begin{align}
p(\boldsymbol{x}, 0)& =e^{-\beta V (\boldsymbol{x}, \lambda_0)+ \beta F(\lambda_0)}, \\
p^*(\boldsymbol{x}, 0) 
& =e^{-\beta V (\boldsymbol{x}, \lambda (\tau))+\beta F(\lambda_\tau)},
\end{align}
\end{subequations}
where $F (\lambda) = - T \log \int_\xv e^{- \beta V(\xv, \lambda)}$ is the equilibrium free energy, a fluctuation theorem can be derived for  
\ba
\Sigma^{\rm cm}[ \bm \gamma]  \equiv 
- \log \frac{p^*(\xv(\tau), 0)}{p(\xv(0), 0)}
- \beta \mathscr Q^{\rm cm}[ \bm \gamma],
\label{Sigma-gamma-1-cm}
\ea
using the standard method of stochastic thermodynamics.  Taking advantage of the first law 
\ba
 \mathscr{W}^{\rm cm}  [ \bm \gamma] 
 + \mathscr{Q}^{\rm cm} [ \bm \gamma]  
 = \Delta V  [ \bm \gamma] ,
 \ea
 one can then prove the following identities:
\begin{align}
 \log \frac{p[ \bm \gamma]}{p^*\left[\hat {\bm \gamma}\right]}
 =\Sigma^{\rm cm}[ \bm \gamma] 
= \beta \mathscr{W}^{\rm cm} [ \bm \gamma]-\beta \Delta F,
\end{align}
where $\Delta F = F(\lambda_\tau) -  F(\lambda_0) $ is the equilibrium free energy difference between the final state and the initial state.  This allows us to express the fluctuation theorem solely in terms of integrated work:
\begin{align}
    {p (\mathscr W^{\rm cm} )}
    =e^{\beta (\mathscr W^{\rm cm} - \Delta F ) } {p^{*}(-\mathscr W^{\rm cm} )}.
    \label{FT-work-cm}
\end{align}

Let us now comment on the differences between our theory and the theory of co-moving frame.  Firstly, the entropy production in the theory of co-moving frame is positive definition only for incompressible fluids, whereas that in our theory is positive definite for arbitrary fluids.  Also, unlike the EP in our theory, the  EP (\ref{dS-tot-cm-1}) in the theory of co-moving frame cannot be decomposed into a positive housekeeping part and a positive excess part.  This also implies that, with heat defined as Eq.~(\ref{shear-flow-dq}), there can be no separate fluctuation theorems for housekeeping  EP  and for excess  EP  in the theory of co-moving frame.  Secondly, the fluctuation theorem (\ref{FT-work-cm}) derived in the theory of co-moving frame applies only to processes starting from equilibrium states, whereas the fluctuation theorems in our theory apply to all processes starting from non-equilibrium states, which include equilibrium states as a special case.  Thirdly, the flow field of the fluid plays a very different role in the two theories.  Whereas in our theory, the term $\gamma \bm v dt$ is treated as a non-conservative driving force, treated separately from friction and external confining potential, in the theory of co-moving frame, this term is treated as an inseparable part of friction force.   For uniform flow with a constant velocity field, it is clearly more natural to describe the physics in the co-moving frame.  For flow fields with shear, however, the co-moving frame is not a Galilean frame, and it is not obvious which theory is conceptually more appealing.  Finally, by comparing Eqs.~(\ref{dft-def1}) with (\ref{cond-generalized-DB-1}), we see that the difference between the two theories may be understood as the difference in the definition of time reversal of non-equilibrium processes.   The system we study in the present work is an example of systems embedded in dissipative backgrounds.  For these systems, there is no unique way of defining the time reversal of dynamic processes.  This results in an ambiguity in the definition of heat, and hence also in the definition of  EP.  Different definitions yield different theories of stochastic thermodynamics. 


\section{Numerical Simulations}
\label{sec:simulation}

In this section, we simulate all four processes as defined in Sec.~\ref{sec:4-processes}, and  and verify all fluctuation theorems (\ref{FT-all}).   To the best of our knowledge, except for a few partial results~\cite{trepagnier_2004_experimental,yoo_2017_molecular}, there has been no systematic verification of fluctuation theorems for housekeeping work and excess work in systems without instantaneous detailed balance.  

\vspace{-3mm}

\subsection{Computing $U$ and $\varphi$}
\label{sec:U-phi}
\vspace{-2mm}

To construct various processes defined in Sec.~\ref{sec:4-processes}, we need $U, \bm\varphi$.  If $\epsilon \ll 1$, they are approximately given by Eqs.~(\ref{Gibbs-parameters}).  If $\epsilon$ is not small, we need to solve the Gibbs gauge condition Eq.~(\ref{Gibbs-gauge-cond-1}) numerically to find $\psi$ and use it in Eqs.~(\ref{Gibbs-gauge-1}) to find $U, \bm \varphi$. The numerical method is explained in App.~\ref{sec:Num-Adj}.  
 
\begin{figure}[h!]
    \centering
    \includegraphics[width=3.2in]{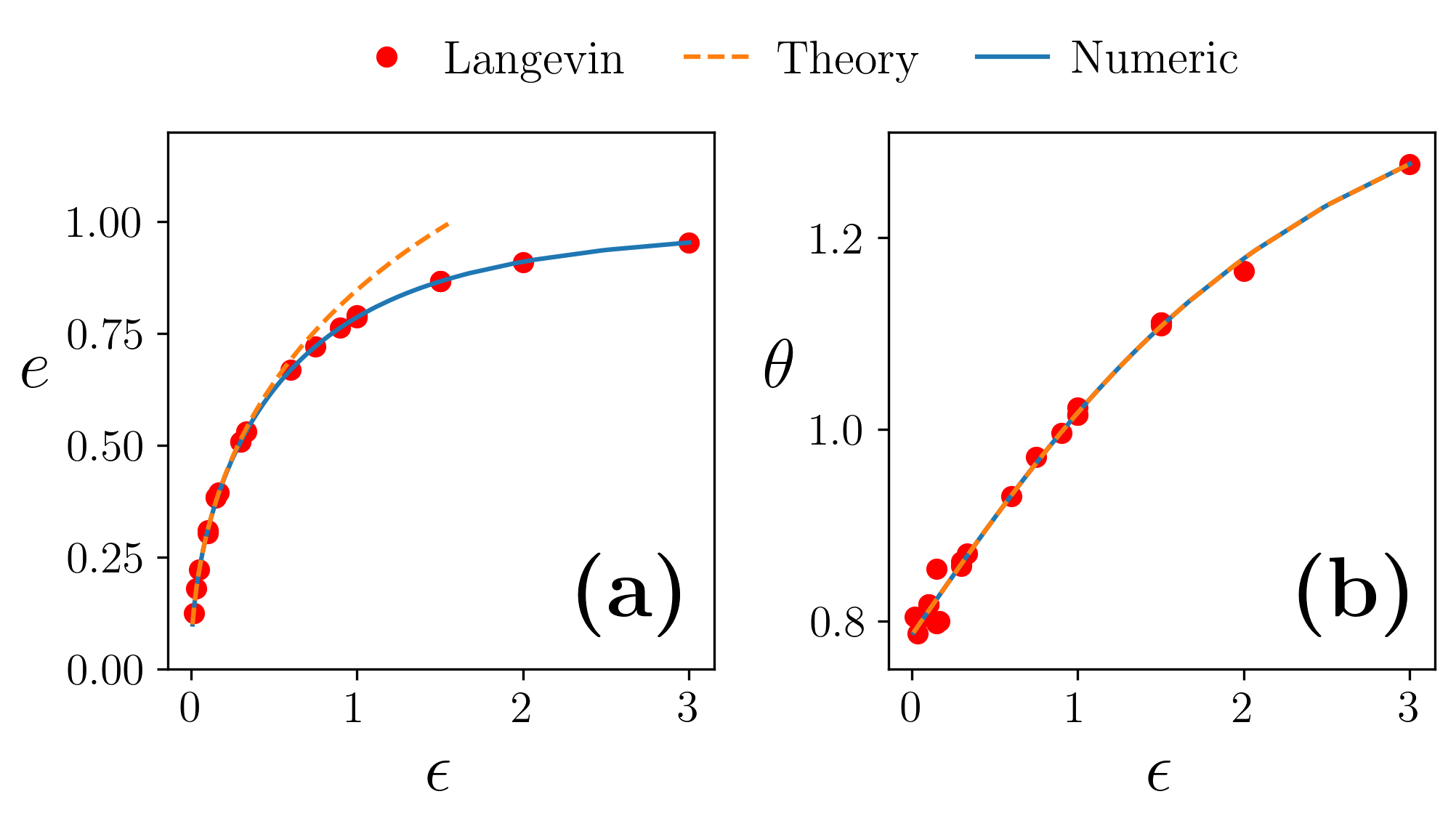}
\vspace{-6mm}
    \caption{ (a) The eccentricity $e$ and  (b) the inclination angle $\theta$ of the contour ellipse of the generalized potential $U$. The red dots are obtained by simulating Langevin Dynamics.  Dashed lines are the analytical result Eqs.~(\ref{Gibbs-parameters}).     Solid lines are obtained by numerically solving the Gibbs gauge condition Eq.~(\ref{Gibbs-gauge-cond-1}). 
    }
    \label{fig::ecc_theta_epsilon}
\end{figure}

To test the accuracy of this method, we calculate a contour line (an ellipse) of thus computed $U$, and plot its eccentricity $e$ and inclination angle $\theta$, i.e., the angle between the major axis and the y-axis. The results are shown in Fig.~\ref{fig::ecc_theta_epsilon} as the solid lines (Numeric).  Also shown there are the corresponding results computed using direct simulation of the Langevin dynamics Eq.~(\ref{langevin-1}) (red dots, Langevin), as well as the analytical results given by Eqs.~(\ref{Gibbs-parameters}) (dashed lines, Theory).  As one can see there, the numeric results agree with the Langevin results for all values of $\epsilon$, which establishes the accuracy of the methods presented  in App.~\ref{sec:Num-Adj} for computation of $U$ and $\bm \varphi$.  By contrast, the analytical results are accurate only for small value of $\epsilon$. 

More numerical testings of our computation methods are supplied in App.~\ref{sec:Num-Adj}. 

\vspace{-2mm}
\subsection{Verification of FTs}

To verify FTs (\ref{FT-all}), we numerically simulate each of the four processes defined in Sec.~\ref{sec:4-processes}.  We generate a large number of trajectories for each process, using the recipe discussed in App.~\ref{appendix::Integration}, compute the total work, the housekeeping work, and the excess work for each trajectory in each process, and thereby obtain the distributions of these works.  The numerical method for computation of work at the trajectory level is explained in Appendix.~\ref{appendix::Numerical calculation of Work}.   In all simulations discussed here, $\epsilon = 1$.  More simulations with different values of $\epsilon$ are presented in App.~\ref{appendix::FT with Other parameter}.


We first verify  Eq.~(\ref{equ::FT of tot work}), which may be rewritten as
\ba
\log \frac{p_{\rm F}(\mathscr W)}{p_{\rm B}(- \mathscr W)} 
= \beta \mathscr W.
\label{equ::FT of tot work-log}
\ea

\begin{table}[t!]
    \centering
    \setlength{\tabcolsep}{3pt}
    \renewcommand{\arraystretch}{1.25}
    \begin{tabular}{|c|c|c|c|c|c|c|}
        \hline
        \multicolumn{1}{|l|}{\multirow{2}{*}{process}} & \multicolumn{2}{c|}{control parameters} & \multirow{2}{*}{\begin{tabular}[c]{@{}c@{}}duration\\ $\tau$\end{tabular} } \\ \cline{2-3}
        &   $K$                  & $x_0,y_0$       &  \\ \hline
        (a)  &   $0.01$                       & $-10 + 20 \frac{t}{\tau}$ &  200, \,100,\, 10, \,1       
        \\ \hline
        (b)  &   $0.01 + 0.02\frac{t}{\tau}$            & 0           & 200, \,100,\, 10, \,1   
        \\ \hline
        (c)  &   $ 0.03 - 0.02 |\frac{2t-\tau}{\tau} | $ & 0           &   200, \,100,\, 10, \,1    \\ \hline
    \end{tabular}
    \caption{  Protocols simulated for verifications of FTs for $ \mathscr W$ and $\mathscr W^{\rm ex}$.  $T=1, \gamma=1, {\zeta}=0.01$. }
    \label{table::1028-ft_wex_combined_TR}
\vspace{-3mm}
\end{table}


\begin{figure}[h!]
    \centering
   \includegraphics[width=3.4in]{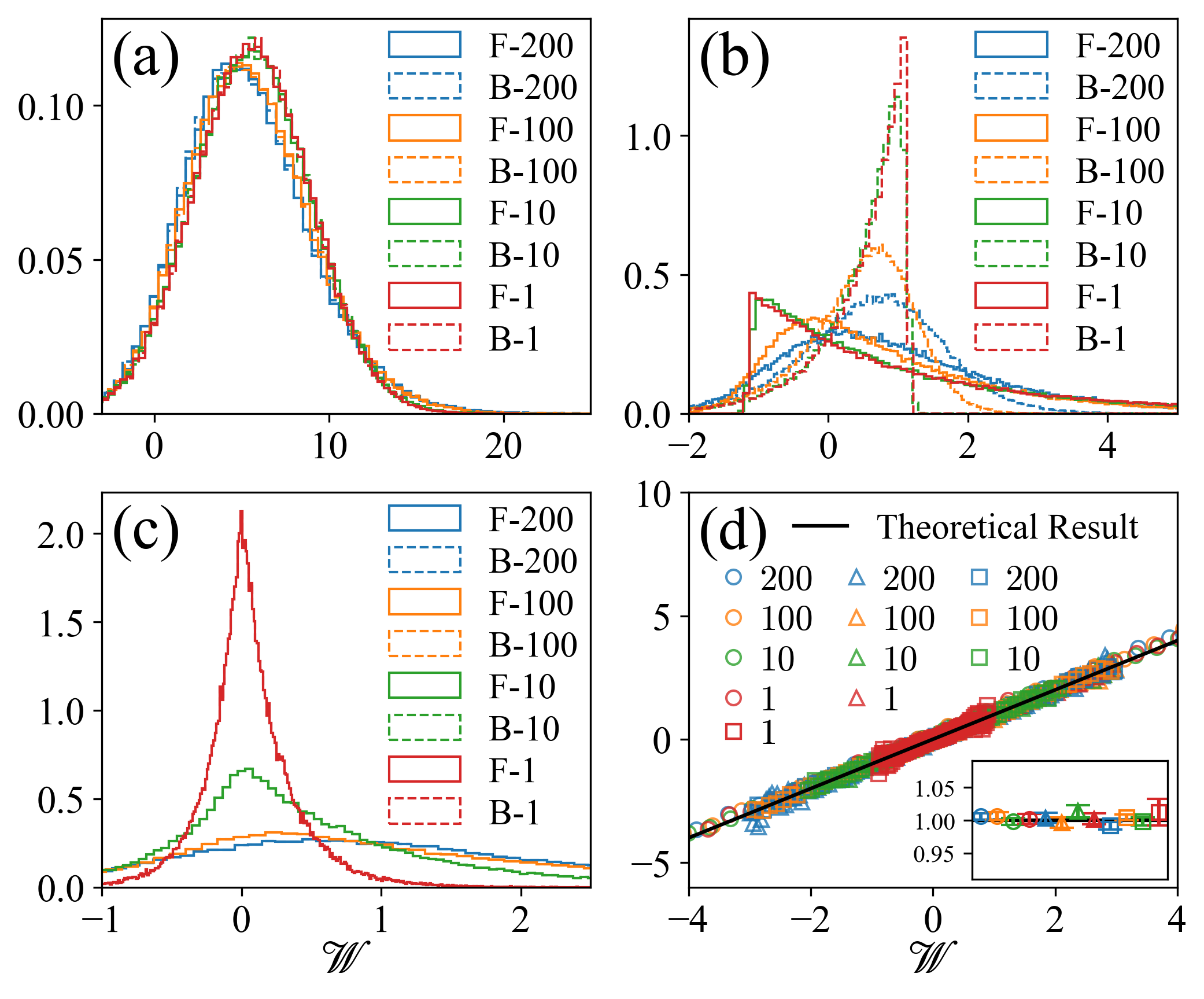}
    \caption{Verification of FT (\ref{equ::FT of tot work-log}). (a), (b), (c): Histograms of the total work $\mathscr{W}^{}$, where all processes are defined in Table \ref{table::1028-ft_wex_combined_TR}. 
    In all legends F, B mean forward and backward respectively. Numbers are durations $\tau$.  (d): Verification of Eq.~(\ref{equ::FT of tot work-log}), where the vertical axis is $\log p_{\rm F} (\mathscr{W})/p_{\rm B}(-\mathscr{W})$. The black straight-line is Eq.~(\ref{equ::FT of tot work-log}).   Circles, triangles, and squares are respectively data from panels (a), (b), (c), whereas numbers are durations of processes.  Inset: The fitting slopes and error bars for each process. }
    \label{fig::1028-ft_w2_combined_TR}
\end{figure}

We simulate three processes that are shown in Table \ref{table::1028-ft_wex_combined_TR}.  The duration $\tau$ of each process is varied systematically, as shown in the last column of the table.  For each protocol, we sample $10^5$ trajectories and compute the distribution of work. We then simulate the backward process, and compute the corresponding distribution of work. These work distributions are displayed in Fig.~\ref{fig::1028-ft_w2_combined_TR} (a), (b), and  (c). Finally we use these distributions to verify the FT (\ref{equ::FT of tot work-log}).  As shown in Fig.~\ref{fig::1028-ft_w2_combined_TR} (d), all data collapse to the black straight-line as predicted by our theory.


Now we verify Eq.~(\ref{equ::FT of hk work}), which may be rewritten as
\ba
\log \frac{p_{\rm F}(\mathscr W^{\rm hk})}{p_{\rm Ad}(- \mathscr W^{\rm hk})} 
= \beta \mathscr W^{\rm hk}.
\label{equ::FT of hk work-log}
\ea

\begin{figure}[t!]
    \centering
 \vspace{-2mm}
   \includegraphics[width=3.4in]{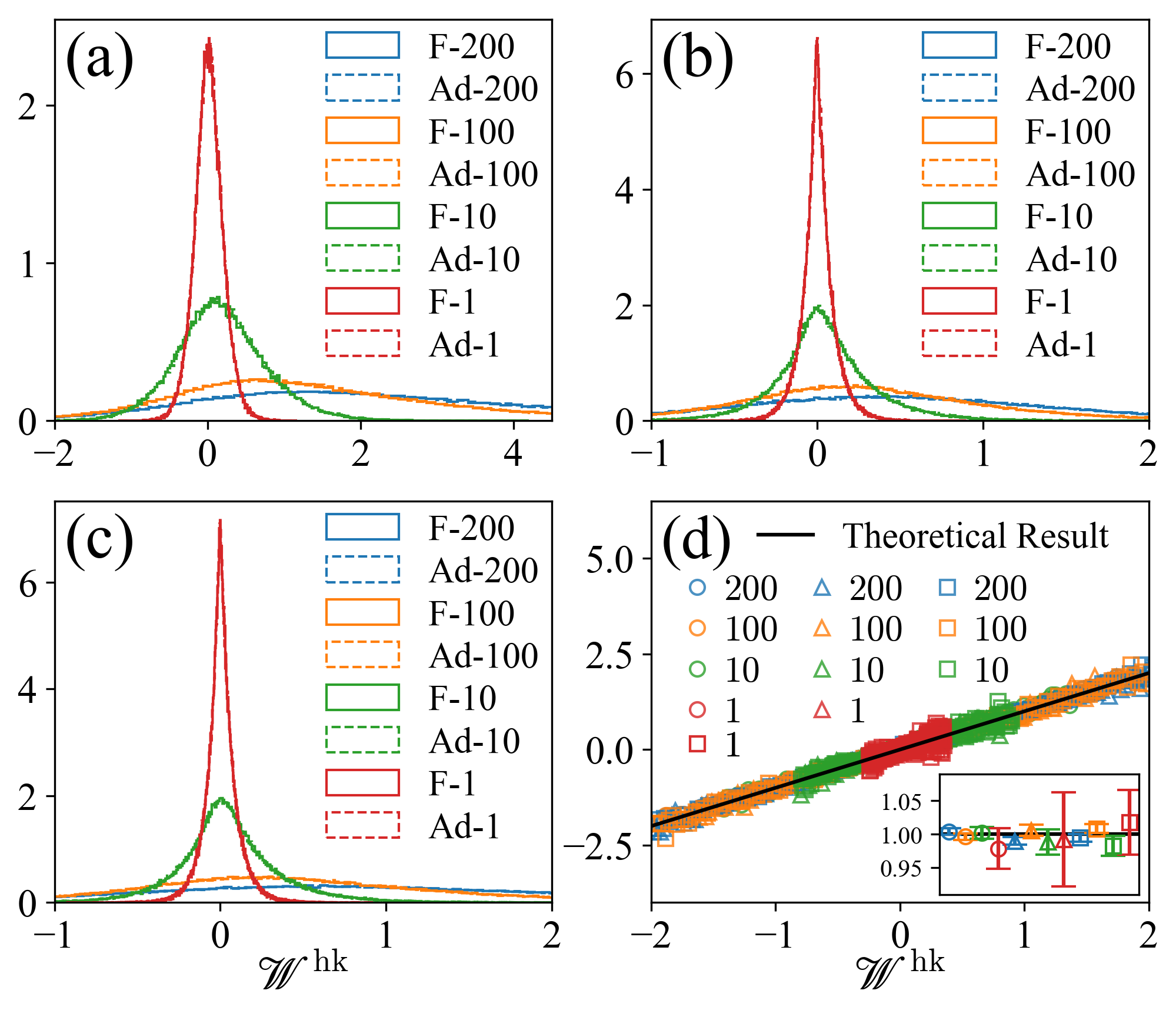}
    \caption{Verification of FT (\ref{equ::FT of hk work-log}). (a), (b), (c): Histograms of the housekeeping work $\mathscr{W}^{\rm hk}$, where all processes are defined in Table \ref{table::1028-ft_whk_combined_TR}. 
    In all legends F, Ad mean forward and backward respectively. Numbers are durations $\tau$.  (d): Verification of FT, where the vertical axis is $\log p_{\rm F} (\mathscr{W}^{\rm hk})/p_{\rm Ad}(-\mathscr{W}^{\rm hk})$. The black straight-line is the FT (\ref{equ::FT of hk work-log}). 
    Circles, triangles, and squares are respectively data from panels (a), (b), (c), whereas numbers are durations of processes.  Inset: The fitting slopes and error bars for each process.}
    \label{fig::1028-ft_whk_combined_TR}
\end{figure}

We simulate three types of forward processes that are shown in Table \ref{table::1028-ft_whk_combined_TR}.   The duration $\tau$ of each process is varied systematically, as shown in the last column of the table.  For each protocol, we sample $10^5$ trajectories and compute the distribution of work. We then do the same for the adjoint processes.  All work distributions are displayed in Fig.~\ref{fig::1028-ft_whk_combined_TR} (a), (b), and (c). Finally, we use these distributions to verify the FT  (\ref{equ::FT of hk work-log}).  As shown in Fig.~\ref{fig::1028-ft_whk_combined_TR} (d), all data collapse to the black straight-line as predicted by our theory.  

\begin{table}[t!]
    \centering
 \vspace{3mm}
    \setlength{\tabcolsep}{3pt}
    \renewcommand{\arraystretch}{1.25}
    \begin{tabular}{|c|c|c|c|c|c|c|}
        \hline
        \multicolumn{1}{|l|}{\multirow{2}{*}{process}} & \multicolumn{2}{c|}{control parameters} & \multirow{2}{*}{\begin{tabular}[c]{@{}c@{}}duration\\ $\tau$\end{tabular} } \\ \cline{2-3}
        &   $K$                  & $x_0,y_0$       &  \\ \hline
        (a)  &   0.01                       & $ 20 -25 |\frac{2t-\tau}{\tau}| $ & 200, \,100,\, 10, \,1      \\ \hline
        (b)  &   $ 0.03 - 0.02 |\frac{2t-\tau}{\tau}| $   & 0           & 200, \,100,\, 10, \,1      \\ \hline
        (c)  &   $ 0.01 $   & 0           & 200, \,100,\, 10, \,1      \\ \hline
    \end{tabular}
        \caption{ All protocols for verifications of FTs for the housekeeping work.  The other parameters are all fixed $T=1, \gamma=1, {\zeta}=0.01$. }
    \label{table::1028-ft_whk_combined_TR}
\end{table}

\begin{figure}[t!]
    \centering
    \includegraphics[width=3.4in]{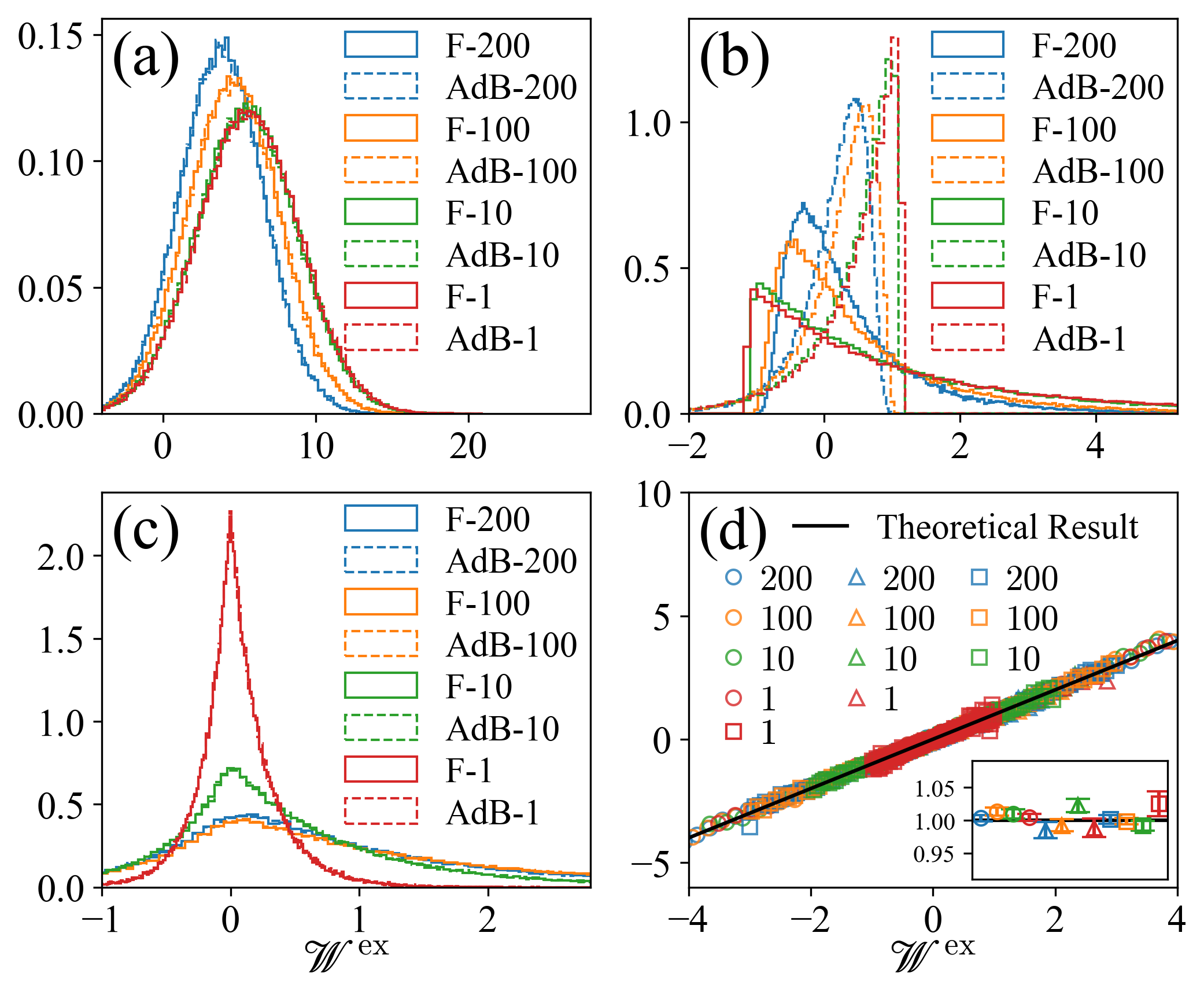}
    \caption{Verification of FT (\ref{equ::FT of ex work-log}) for the excess work. (a), (b), (c): Histograms of the excess work $\mathscr{W}^{\rm ex}$, where all processes are defined in Table \ref{table::1028-ft_wex_combined_TR}. 
    In all legends F, AdB mean forward and backward respectively and numbers are durations $\tau$.  (d): Verification of FT, where the vertical axis is $\log p_{\rm F} (\mathscr{W}^{\rm ex})/p_{\rm AdB}(-\mathscr{W}^{\rm ex})$. The black straight-line is the FT (\ref{equ::FT of ex work-log}). 
    Circles, triangles, and squares are respectively data from panels (a), (b), (c), whereas numbers are durations of processes.  Inset: The fitting slopes and error bars for each process. }
    \label{fig::1028-ft_wex_combined_TR}
\vspace{-3mm}
\end{figure}


Now we verify Eq.~(\ref{equ::FT of ex work}), which may be rewritten as
\ba
\log \frac{p_{\rm F}(\mathscr W^{\rm ex})}{p_{\rm AdB}(- \mathscr W^{\rm ex})} 
= \beta \mathscr W^{\rm ex}.
\label{equ::FT of ex work-log}
\ea

We simulate the same processes as shown in Table \ref{table::1028-ft_wex_combined_TR}, and compute the distributions of excess work.  We then do the same for the adjoint backward processes. All work distributions are displayed in Fig.~\ref{fig::1028-ft_wex_combined_TR} (a), (b), and (c). Finally, we use these distributions to verify the FT  (\ref{equ::FT of ex work-log}).  As shown in Fig.~\ref{fig::1028-ft_wex_combined_TR} (d), all data collapse to the black straight-line as predicted by our theory.

\section{Conclusion}
\label{sec:conclusion}
In this work, we have developed a theory of stochastic thermodynamics for over-damped Brownian motion in a flowing fluid.  To the best of our knowledge, this is the first concrete example of non-equilibrium small systems for which fluctuation theorems of the total work, the housekeeping work, and the excess work are explicitly established and verified.  The analytic and numerical methods we employed here should be valuable for study of other non-equilibrium systems.  

The authors acknowledge support from NSFC via grant \#12375035, as well as Shanghai Municipal Science and Technology Major Project (Grant No.2019SHZDZX01).


\appendix
\onecolumngrid


\section{The Numerical Methods}\label{appendix::transformation form V to U}
\subsection{Computation of $U$ and $ \varphi$}
\label{sec:Num-Adj}

%

Here we explain how to compute the coefficients $A, B, C, D, E$ in the expansion Eq.~(\ref{equ::U in ABC form-0}).  We consider slightly more general forms of quadratic confining potential and linear incompressible velocity field:
\ba
    U^{0} &=& \beta V 
    = a_0 x^2 + b_0 x y + c_0 y^2 + d_0 x + e_0 y + f_0, 
 \label{U0-App}   \\
\bm \varphi^0 &=& \beta \gamma\,  \bm v 
=   \beta \gamma\,  ( {y \zeta_x} \, \hat{\bm e}_x 
+ {x \zeta_y}  \, \hat{\bm e}_y ).
 \label{varphi0-App} 
\ea

Using Eq.~(\ref{gauge-transform}), we may rewrite the Gibbs gauge condition (\ref{Gibbs-gauge-cond-1}) as
\ba
\partial_i ( \varphi_i^0  + \partial_i (U - U^0)  ) 
- ( \varphi_i^0  + \partial_i  (U - U^0) ) \partial_i   U  = 0.
     \label{Gibbs-gauge-cond-App}
\ea
Note that the l.h.s. is also a quadratic form of $\xv$. 

We insert Eqs.~(\ref{U0-App}), (\ref{varphi0-App}), and (\ref{equ::U in ABC form-0}) into Eq.~(\ref{Gibbs-gauge-cond-App}) and compare all coefficients of the quadratic form, we find following set of nonlinear equations:
\begin{subequations}
\begin{align}
    x^2 :  \quad & 4A(a_0-A) + B(b_0-B -{\beta \gamma \zeta_y}) =0,  \\
    y^2 : \quad& 4C(c_0-C) + B(b_0-B -{\beta \gamma \zeta_x}) =0,  \\
    xy : \quad& B(a_0-A) + A(b_0-B -{\beta \gamma \zeta_x}) + B(c_0-C) + C(b_0-B -{\beta \gamma \zeta_y}) = 0,  \\
    x : \quad& 2D(a_0-A) + 2A(d_0-D) + E(b_0-B -{\beta \gamma \zeta_y}) + B(e_0-E) = 0,  \\
    y : \quad& D(b_0-B -{\beta \gamma \zeta_x}) + B(d_0-D) + 2E(c_0-C) + 2C(e_0-E) = 0,  \\
    o(1) : \quad& 2(a_0-A)+ 2(c_0-C) - D(d_0-D) - E(e_0-E) = 0 .
\end{align}\label{equ::FP-algebra equation}
\end{subequations}
Notice that only five of these equations are independent, since there are only five unknowns $A, B, C, D, E$ appearing in these equations. 

\subsection{Testing of numerical methods}

\begin{figure}[h!]
    \centering
    \includegraphics[width=4 in]{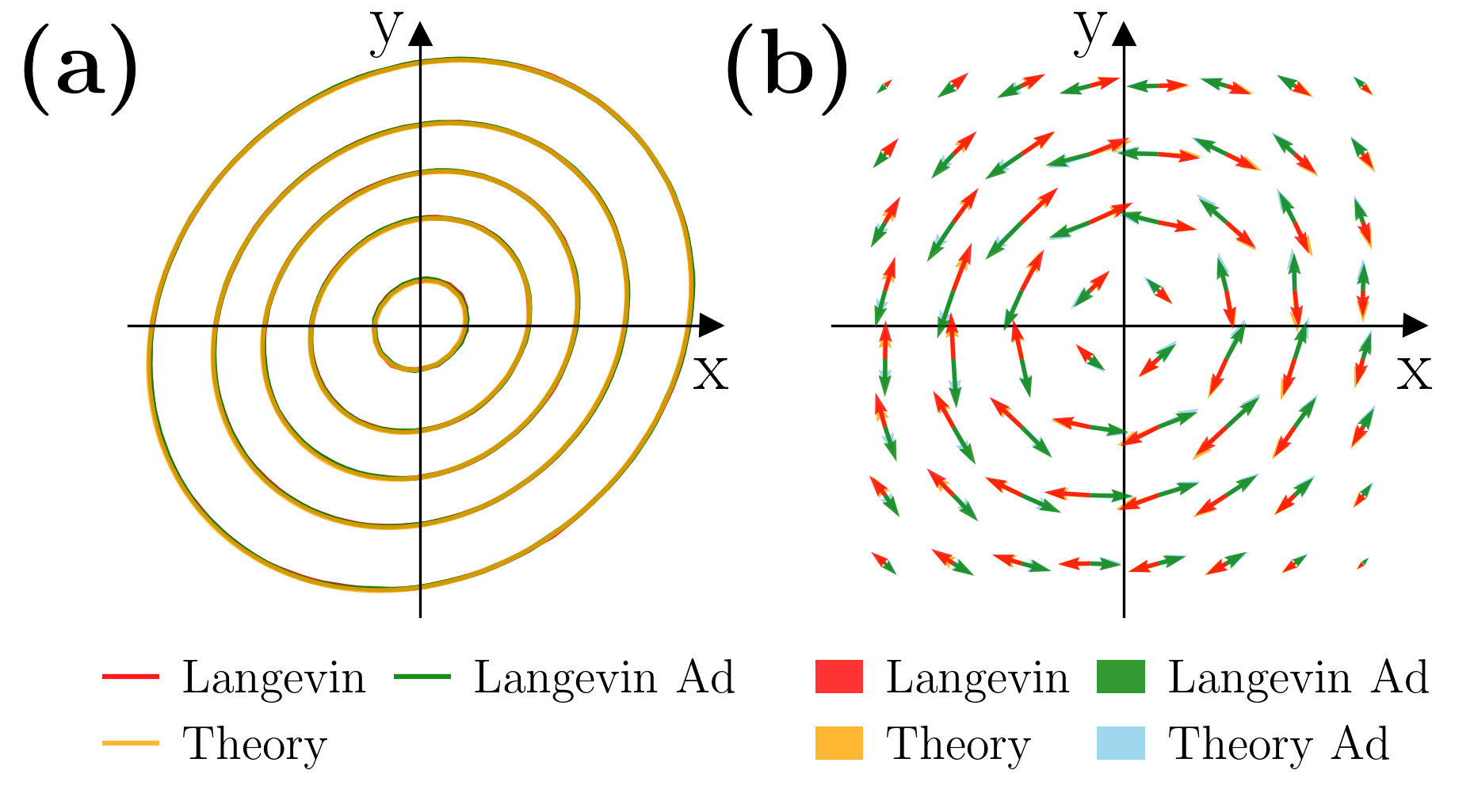}
    \caption{(a): Contour plots of NESS probability density function (PDF).  
      (b):  NESS probability currents. 
     Relevant parameters: ${\zeta}=0.01$, $K=0.01$, $x_0= y_0 = 0$, $T=1$, $\gamma=0.3$, and $\epsilon = 0.3$.}
    \label{fig::ug_nc_force_exp_gd001_k001_x0_g03}
\end{figure}

Here we supply more testing of the analytic results (\ref{Gibbs-parameters}) as well as the numerical results for $U, \bm \varphi$, obtained using the method discussed in Eq.~(\ref{sec:Num-Adj}). 

We simulate the Langevin dynamics (\ref{langevin-1}) with the confine potential and flow field given by Eqs.~(\ref{V-v-1}), and compute the NESS pdf and probability current.  We do the same thing for the adjoint dynamics, where the confining potential and the flow field are given by Eqs.~(\ref{V-v-Ad}). 

Firstly we let $\epsilon = 0.3$, so that the analytical results (\ref{Gibbs-parameters}) are expected to be good. 

In Fig.~\ref{fig::ug_nc_force_exp_gd001_k001_x0_g03}(a)  we plot the contour lines of the NESS pdfs both for the original dynamics and the adjoint dynamics, computed using simulation data.  In the same figure we also show the contour lines of the NESS pdf given by analytic results, i.e., Eqs.~(\ref{p^ss-def-1}) and (\ref{Gibbs-parameters}).  As one can see there, all results agree with each other up to high precision.   
 
 In Fig.~\ref{fig::ug_nc_force_exp_gd001_k001_x0_g03}(b), we plot the NESS probability currents of both the original dynamics and the adjoint dynamics.  As one can see, the probability current of the forward process is the opposite of that of the adjoint process.  Additionally, theoretical results agree with numerical results.

Now we let $\epsilon = 1$, so that the analytical results (\ref{Gibbs-parameters}) are not expected to be good. We will then use the numerical method discussed in App.~\ref{sec:Num-Adj} to compute $U, \bm \varphi$. 
 
\begin{figure}[t!]
    \centering
    \includegraphics[width=4 in]{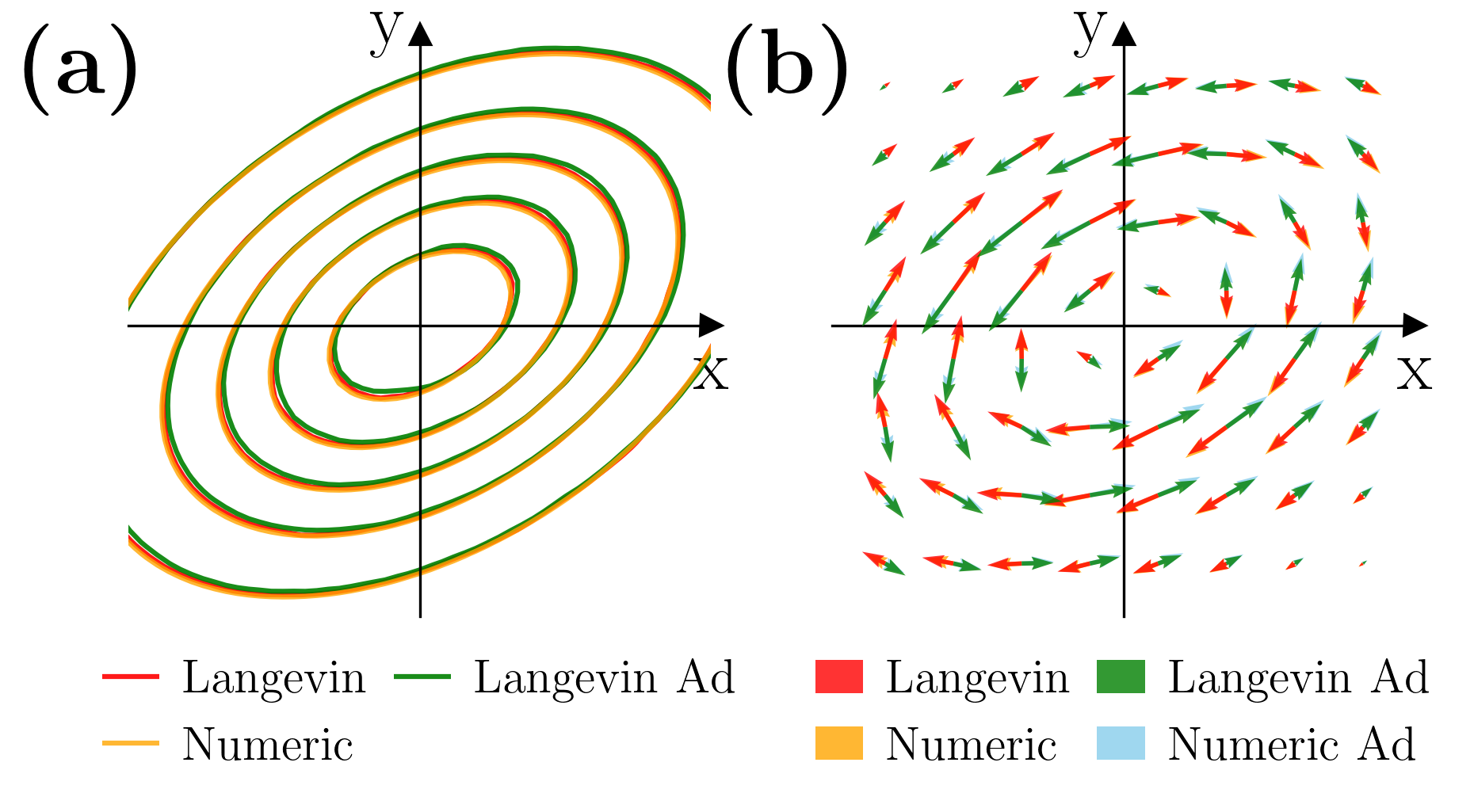}
     \caption{(a): Contour plots of NESS probability density function (PDF).  
     (b):  NESS probability currents. 
     Relevant parameters: ${\zeta}=0.01$, $K=0.01$, $x_0= y_0 = 0$, $T=1$, $\gamma=0.3$, and $\epsilon = 1$.}
    \label{fig::ug_nc_force_gd001_k001_x0}
\vspace{-5mm}
\end{figure}

In Fig.~\ref{fig::ug_nc_force_gd001_k001_x0}(a)  we plot the contour lines of the NESS pdfs both for the original dynamics and the adjoint dynamics, computed using simulation data.  In the same figure we also show the contour lines of the NESS pdf computed using the method discussed in App.~\ref{sec:Num-Adj}.  As one can see there, all results agree with each other up to high precision.   
 
 In Fig.~\ref{fig::ug_nc_force_gd001_k001_x0}(b), we plot the NESS probability currents of both the original dynamics and the adjoint dynamics, computed both using direction simulation of the Langevin dynamics, and using the method discussed in App.~\ref{sec:Num-Adj}.  As one can see, the probability current of the forward process is the opposite of that of the adjoint process.  Additionally, simulation results agree with numerical results.  

\subsection{Numerical integration of Langevin dynamics}
\label{appendix::Integration}

To numerically solve Langevin equation(\ref{langevin-1}), we use the first-order Euler-Maruyama scheme \cite{Kloeden1992}.

First we discretize $t$ with step size $=  0.001$:
\ba
\Delta t &=&t_{n+1} - t_n ,\\
\lambda_n & \equiv & \lambda(t_n),\\
 \Delta \boldsymbol{x}_{n+1}  &\equiv& 
      \boldsymbol{x}(t_{n+1} ) -   \boldsymbol{x}(t_n),
       \label{equ::langevin dynamics numeric}
\ea
so that Eq.~(\ref{langevin-1}) is discretized as follows:
\begin{align}
\Delta \boldsymbol{x}_{n+1}  &=  \left[ 
    \boldsymbol{v}(t_n)
     + \frac{\boldsymbol{F}(t_n)}{\gamma} 
    \right]   \Delta t 
     + \sqrt{\frac{2 T\Delta t}{\gamma}} \boldsymbol{\xi}(t_n), 
     \label{Langevin-discretized}
\end{align}
where $\boldsymbol{\xi} = (\xi_1, \xi_2)$ is a 2d vector of normalized Gaussian random variables, and $\boldsymbol{F}(t_n)$ and $\boldsymbol{v}(t_n)$ are respectively the discretized force and fluid velocity:
\begin{align}
        \boldsymbol{F}(t_n) &= -\partial_x V(\xv(t_n), \lambda_n)  \, \hat{\bm e}_x
         -\partial_y V(\xv(t_n), \lambda_n) \,  \hat{\bm e}_y, \\
    \boldsymbol{v}(t_n) &= \zeta \, y(t_n) \,\hat{\bm e}_x. 
\end{align}
Note that the potential $V(\xv, \lambda)$ and the flow field $\bm v(\xv)$ are given in Eqs.~(\ref{V-v-1}).

We then numerically solve the discretized equations (\ref{Langevin-discretized}).

\subsection{Calculation of Work}\label{appendix::Numerical calculation of Work}

The total work $d\mathscr{W}^{}$ along a trajectory $\bm \gamma$ is given in Eq.~(\ref{W-F-decomp-1}).  It  can be discretized as
\begin{align}
    \mathscr{W}^{} [ \bm \gamma] 
    &= \sum_{n=0}^{n=N} 
    T  (\lambda_{n+1}- \lambda_n )\,  
    \partial_\lambda U  (\xv_n, \lambda_n)
     + \sum_{n=0}^{n=N} 
     T \bm \varphi (\xv_{n+1/2}, \lambda_n) \cdot
     \Delta \xv_{n+1} . \label{equ::discretized tot work}
\end{align}
where $  \Delta \xv_{n+1} $ is defined in Eq.~(\ref{equ::langevin dynamics numeric}), and $U$ is given in Eq.(\ref{equ::U in ABC form-0}), whereas $\xv_{n+1/2}$ is defined as
\begin{gather}
    \xv_{n+{1}/{2}} = \frac{\xv (t_n)+ \xv (t_{n+1})}{2}.
\end{gather}
It is important to evaluate $\bm \varphi$ at $ \xv_{n+{1}/{2}} $ rather than any other place, in order to correctly compute the Stratonovich product in Eq.~(\ref{W-F-decomp-1}).

The housekeeping work and excess work, defined in Eqs.(\ref{W-hk-ex-def}) can be similarly discretized:
\begin{align}
    \mathscr{W}^{\rm hk}[ \bm \gamma]
    &= \sum_{n=0}^{n=N} 
     T \bm \varphi (\xv_{n+1/2}, \lambda_n) \cdot
     \Delta \xv_{n+1} .
     \label{equ::discretized hk work}\\
    \mathscr{W}^{\rm ex}[ \bm \gamma] 
    &=
     \sum_{n=0}^{n=N} 
    T  (\lambda_{n+1}- \lambda_n )\, 
     \partial_\lambda U  (\xv_n, \lambda_n)
     . \label{equ::discretized ex work}
\end{align}

\section{FT with Other parameter}
\label{appendix::FT with Other parameter}
\subsection{Small Shear Rate}
In this part, we also verify FTs (\ref{FT-all}).  In all simulations discussed here, we set parameter $T=1, \gamma=0.3, {\zeta}=0.01$, and $\epsilon = 0.3$.

\begin{figure}[h!]
  \centering
  \includegraphics[width=4in]{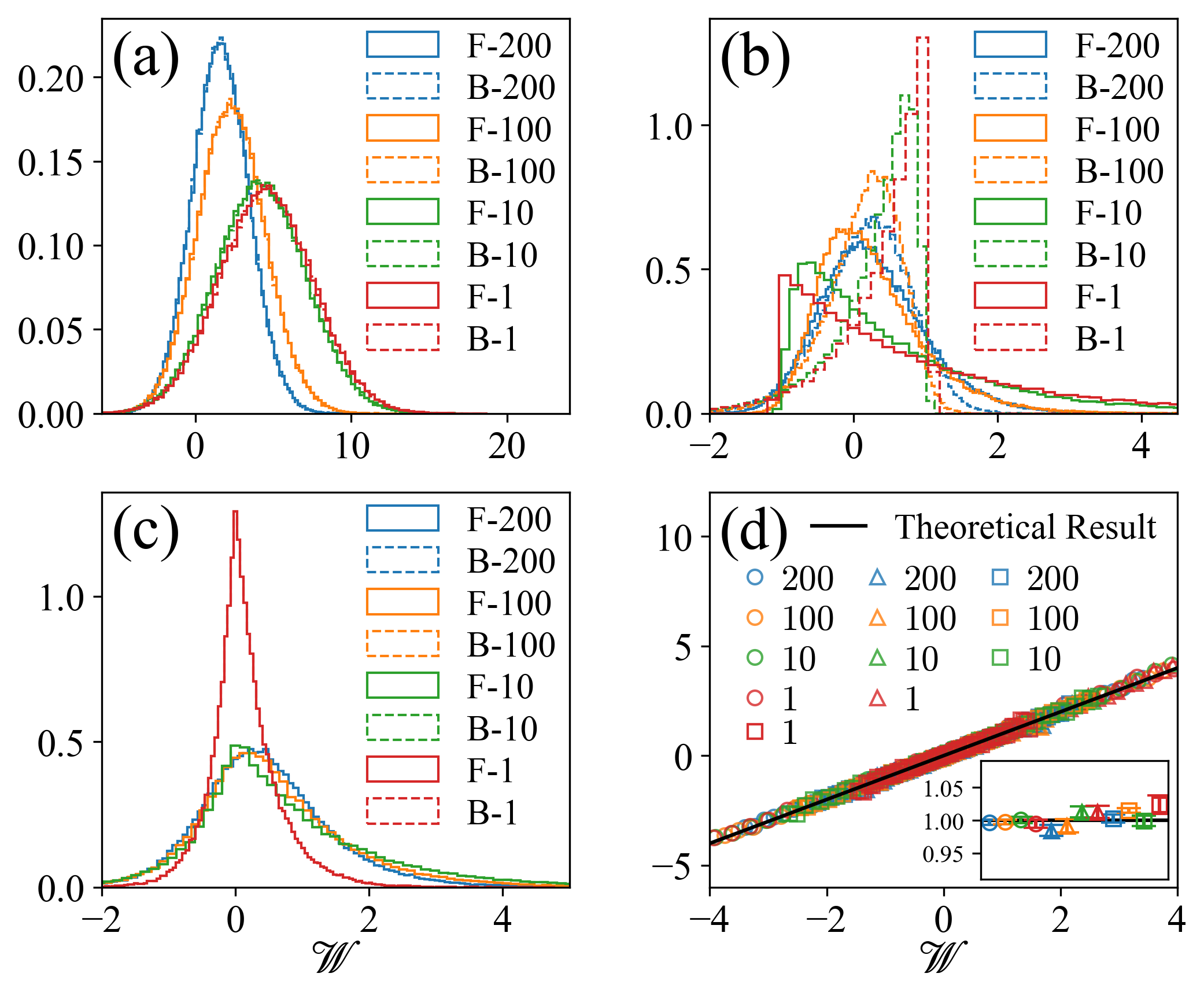}
  \caption{
    Verification of FT (\ref{equ::FT of tot work-log}). (a), (b), (c): Histograms of the total work $\mathscr{W}^{}$, where all processes are defined in Table \ref{table::1028-ft_wex_combined_TR}. 
    In all legends F, B mean forward and backward respectively. Numbers are durations $\tau$.  (d): Verification of Eq.~(\ref{equ::FT of tot work-log}), where the vertical axis is $\log p_{\rm F} (\mathscr{W})/p_{\rm B}(-\mathscr{W})$. The black straight-line is Eq.~(\ref{equ::FT of tot work-log}).   Circles, triangles, and squares are respectively data from panels (a), (b), (c), whereas numbers are durations of processes.  Inset: The fitting slopes and error bars for each process.
  }
  \label{fig::240226-t7920-ft_w2_combined_TR}
\end{figure}

We first verify  Eq.~(\ref{equ::FT of tot work}), which may be rewritten as in Eq.~(\ref{equ::FT of tot work-log}).  We simulate three processes that are shown in Table \ref{table::1028-ft_wex_combined_TR}.  The duration $\tau$ of each process is varied systematically, as shown in the last column of the table.  For each protocol, we sample $10^5$ trajectories and compute the distribution of work. We then simulate the backward process, and compute the corresponding distribution of work. These work distributions are displayed in Fig.~\ref{fig::240226-t7920-ft_w2_combined_TR} (a), (b), and  (c). Finally we use these distributions to verify the FT (\ref{equ::FT of tot work-log}).  As shown in Fig.~\ref{fig::240226-t7920-ft_w2_combined_TR} (d), all data collapse to the black straight-line as predicted by our theory.

\begin{figure}[h!]
  \centering
  \setlength{\tabcolsep}{3pt}
  \renewcommand{\arraystretch}{1.25}
  \includegraphics[width=4in]{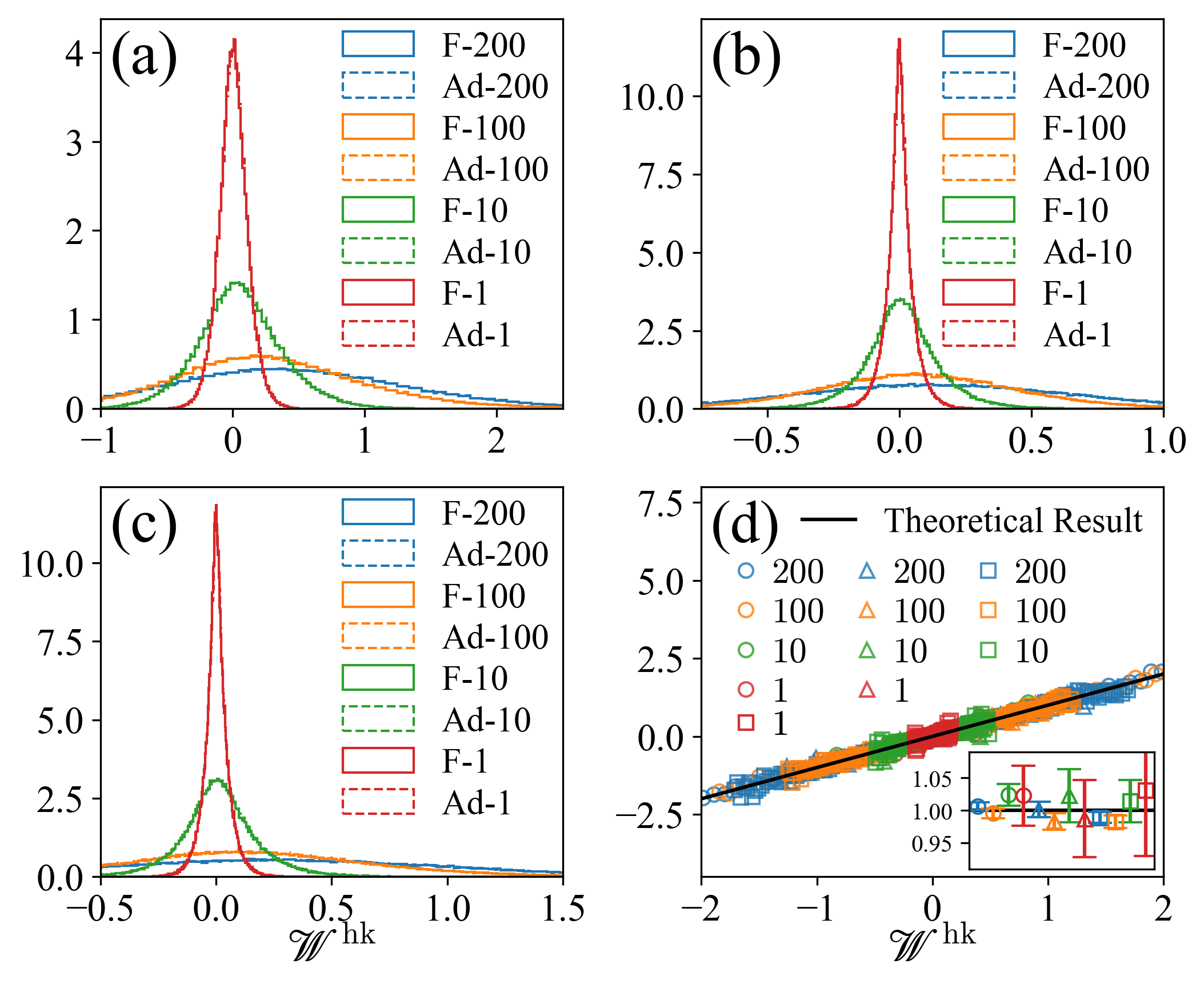}
  \caption{
    Verification of FT (\ref{equ::FT of hk work-log}). (a), (b), (c): Histograms of the housekeeping work $\mathscr{W}^{\rm hk}$, where all processes are defined in Table \ref{table::1028-ft_whk_combined_TR}. 
    In all legends F, Ad mean forward and backward respectively. Numbers are durations $\tau$.  (d): Verification of FT, where the vertical axis is $\log p_{\rm F} (\mathscr{W}^{\rm hk})/p_{\rm Ad}(-\mathscr{W}^{\rm hk})$. The black straight-line is the FT (\ref{equ::FT of hk work-log}). 
    Circles, triangles, and squares are respectively data from panels (a), (b), (c), whereas numbers are durations of processes.  Inset: The fitting slopes and error bars for each process.
  }
\vspace{-3mm}
  \label{fig::240226-t7920-ft_whk_combined_TR}
\end{figure}

Then we verify Eq.~(\ref{equ::FT of hk work}), which may be rewritten as in Eq.~(\ref{equ::FT of hk work-log}). We simulate three types of forward processes that are shown in Table \ref{table::1028-ft_whk_combined_TR}.   The duration $\tau$ of each process is varied systematically, as shown in the last column of the table.  For each protocol, we sample $10^5$ trajectories and compute the distribution of work. We then do the same for the adjoint processes.  All work distributions are displayed in Fig.~\ref{fig::240226-t7920-ft_whk_combined_TR} (a), (b), and (c). Finally, we use these distributions to verify the FT  (\ref{equ::FT of hk work-log}).  As shown in Fig.~\ref{fig::240226-t7920-ft_whk_combined_TR} (d), all data collapse to the black straight-line as predicted by our theory.

\begin{figure}[h!]
  \centering
  \includegraphics[width=4in]{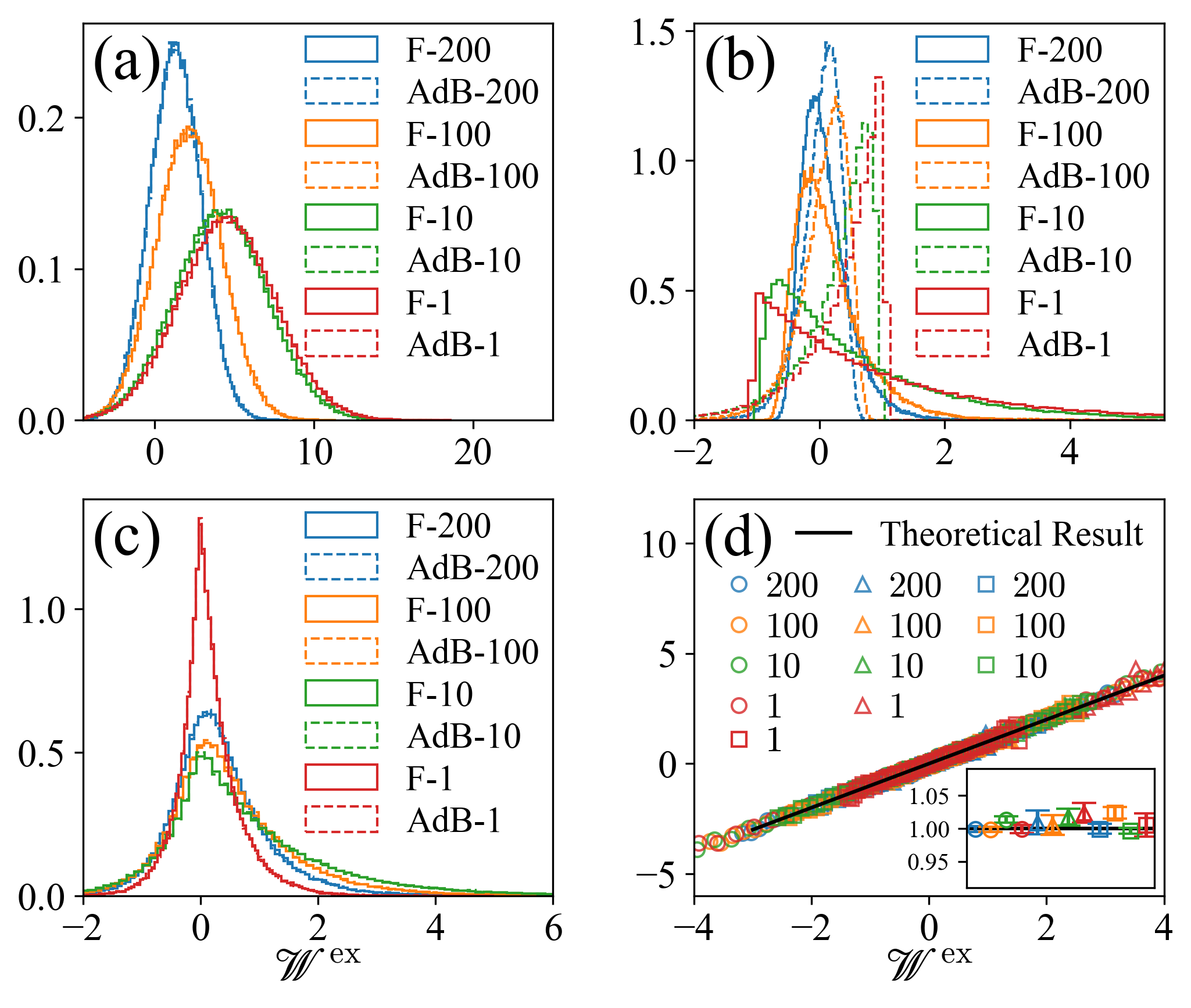}
  \caption{
    Verification of FT (\ref{equ::FT of ex work-log}) for the excess work. (a), (b), (c): Histograms of the excess work $\mathscr{W}^{\rm ex}$, where all processes are defined in Table \ref{table::1028-ft_wex_combined_TR}. 
    In all legends F, AdB mean forward and backward respectively and numbers are durations $\tau$.  (d): Verification of FT, where the vertical axis is $\log p_{\rm F} (\mathscr{W}^{\rm ex})/p_{\rm AdB}(-\mathscr{W}^{\rm ex})$. The black straight-line is the FT (\ref{equ::FT of ex work-log}). 
    Circles, triangles, and squares are respectively data from panels (a), (b), (c), whereas numbers are durations of processes.  Inset: The fitting slopes and error bars for each process.
  }
  \label{fig::240226-t7920-ft_ex_combined_TR}
\end{figure}

Finally we verify Eq.~(\ref{equ::FT of ex work}), which may be rewritten as in Eq.~(\ref{equ::FT of ex work-log}). We simulate the same processes as shown in Table \ref{table::1028-ft_wex_combined_TR}, and compute the distributions of excess work.  We then do the same for the adjoint backward processes. All work distributions are displayed in Fig.~\ref{fig::240226-t7920-ft_ex_combined_TR} (a), (b), and (c). Finally, we use these distributions to verify the FT  (\ref{equ::FT of ex work-log}).  As shown in Fig.~\ref{fig::240226-t7920-ft_ex_combined_TR} (d), all data collapse to the black straight-line as predicted by our theory.  

\subsection{Larger Shear Rate}
In this part, we also verify FTs (\ref{FT-all}).  In all simulations discussed here, we set parameter $T=1, \gamma=1, {\zeta}=0.03$, and $\epsilon = 3$.

\begin{figure}[h!]
  \centering
  \includegraphics[width=4in]{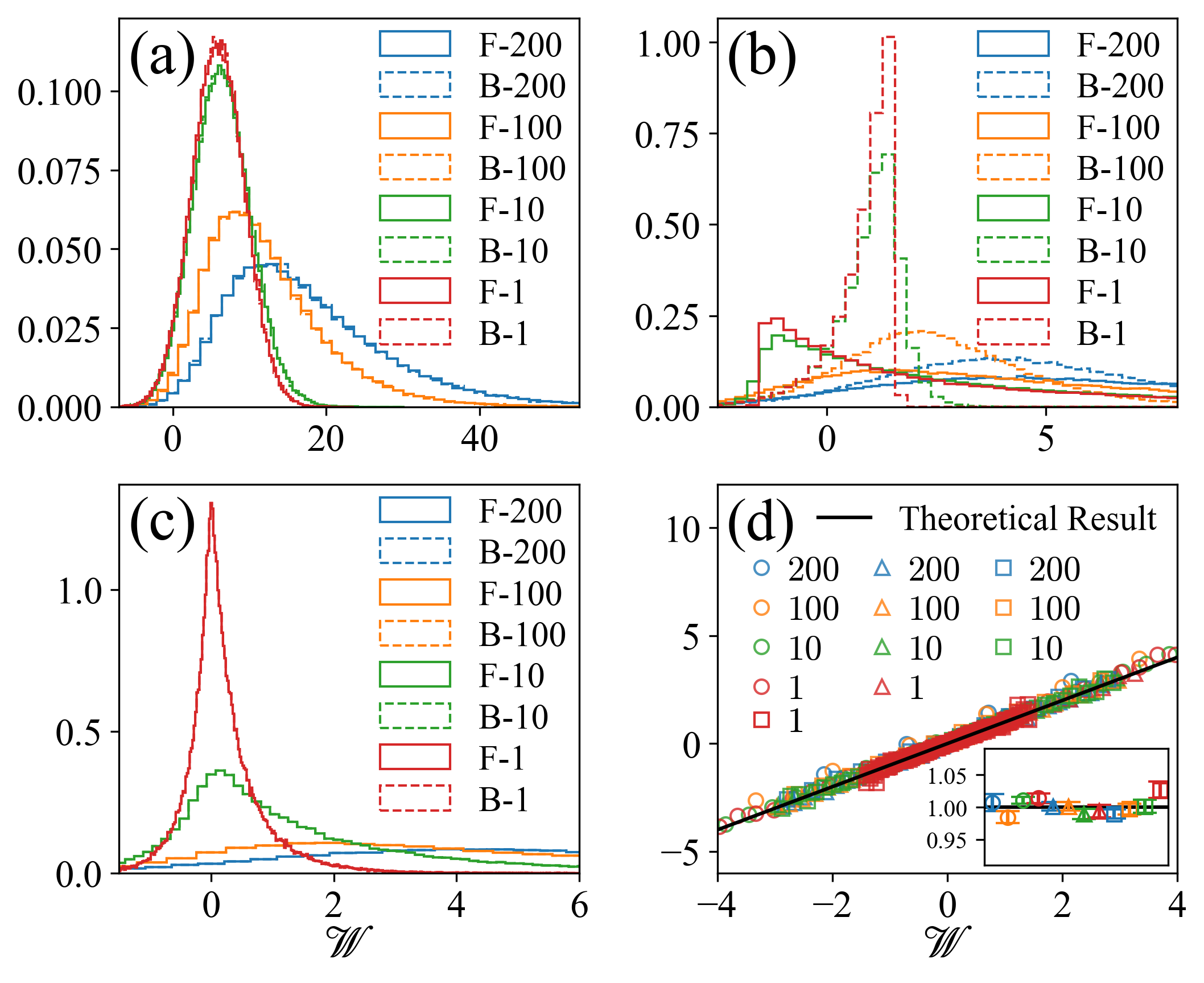}
  \caption{
    Verification of FT (\ref{equ::FT of tot work-log}). (a), (b), (c): Histograms of the total work $\mathscr{W}^{}$, where all processes are defined in Table \ref{table::1028-ft_wex_combined_TR}. 
    In all legends F, B mean forward and backward respectively. Numbers are durations $\tau$.  (d): Verification of Eq.~(\ref{equ::FT of tot work-log}), where the vertical axis is $\log p_{\rm F} (\mathscr{W})/p_{\rm B}(-\mathscr{W})$. The black straight-line is Eq.~(\ref{equ::FT of tot work-log}).   Circles, triangles, and squares are respectively data from panels (a), (b), (c), whereas numbers are durations of processes.  Inset: The fitting slopes and error bars for each process.
  }
\vspace{-3mm}
  \label{fig::240226-gamma1_gd003_ft_w2_combined_TR}
\end{figure}

We first verify  Eq.~(\ref{equ::FT of tot work}), which may be rewritten as in Eq.~(\ref{equ::FT of tot work-log}).  We simulate three processes that are shown in Table \ref{table::1028-ft_wex_combined_TR}.  The duration $\tau$ of each process is varied systematically, as shown in the last column of the table.  For each protocol, we sample $10^5$ trajectories and compute the distribution of work. We then simulate the backward process, and compute the corresponding distribution of work. These work distributions are displayed in Fig.~\ref{fig::240226-gamma1_gd003_ft_w2_combined_TR} (a), (b), and  (c). Finally we use these distributions to verify the FT (\ref{equ::FT of tot work-log}).  As shown in Fig.~\ref{fig::240226-gamma1_gd003_ft_w2_combined_TR} (d), all data collapse to the black straight-line as predicted by our theory.

\begin{figure}[h!]
  \centering
  \setlength{\tabcolsep}{3pt}
  \renewcommand{\arraystretch}{1.25}
  \includegraphics[width=4in]{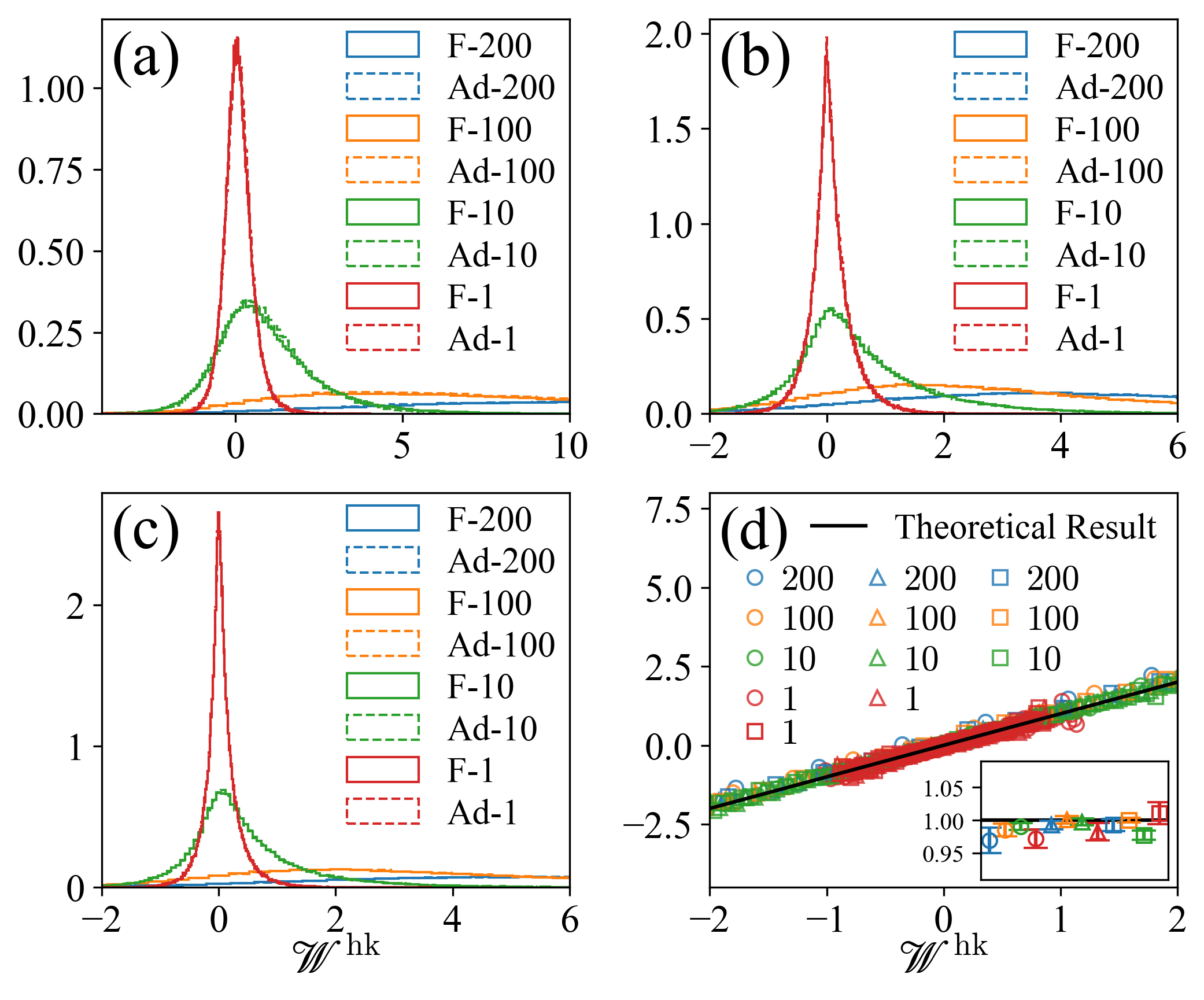}
  \caption{
    Verification of FT (\ref{equ::FT of hk work-log}). (a), (b), (c): Histograms of the housekeeping work $\mathscr{W}^{\rm hk}$, where all processes are defined in Table \ref{table::1028-ft_whk_combined_TR}. 
    In all legends F, Ad mean forward and backward respectively. Numbers are durations $\tau$.  (d): Verification of FT, where the vertical axis is $\log p_{\rm F} (\mathscr{W}^{\rm hk})/p_{\rm Ad}(-\mathscr{W}^{\rm hk})$. The black straight-line is the FT (\ref{equ::FT of hk work-log}). 
    Circles, triangles, and squares are respectively data from panels (a), (b), (c), whereas numbers are durations of processes.  Inset: The fitting slopes and error bars for each process.
  }
  \label{fig::240226-gamma1_gd003_ft_whk_combined_TR}
\end{figure}

Then we verify Eq.~(\ref{equ::FT of hk work}), which may be rewritten as in Eq.~(\ref{equ::FT of hk work-log}). We simulate three types of forward processes that are shown in Table \ref{table::1028-ft_whk_combined_TR}.   The duration $\tau$ of each process is varied systematically, as shown in the last column of the table.  For each protocol, we sample $10^5$ trajectories and compute the distribution of work. We then do the same for the adjoint processes.  All work distributions are displayed in Fig.~\ref{fig::240226-gamma1_gd003_ft_whk_combined_TR} (a), (b), and (c). Finally, we use these distributions to verify the FT  (\ref{equ::FT of hk work-log}).  As shown in Fig.~\ref{fig::240226-gamma1_gd003_ft_whk_combined_TR} (d), all data collapse to the black straight-line as predicted by our theory.

\begin{figure}[h!]
  \centering
  \includegraphics[width=4in]{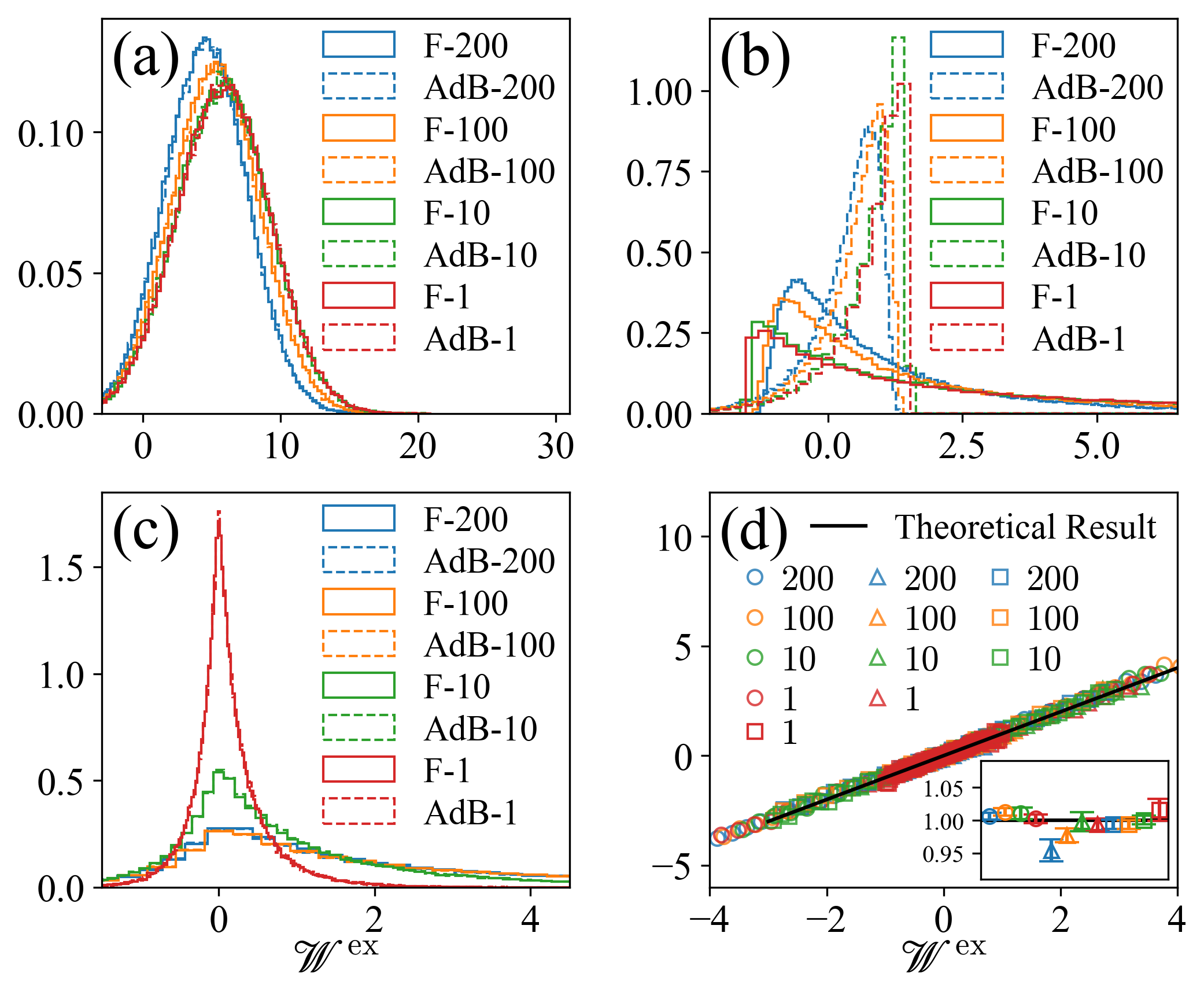}
  \caption{
    Verification of FT (\ref{equ::FT of ex work-log}) for the excess work. (a), (b), (c): Histograms of the excess work $\mathscr{W}^{\rm ex}$, where all processes are defined in Table \ref{table::1028-ft_wex_combined_TR}. 
    In all legends F, AdB mean forward and backward respectively and numbers are durations $\tau$.  (d): Verification of FT, where the vertical axis is $\log p_{\rm F} (\mathscr{W}^{\rm ex})/p_{\rm AdB}(-\mathscr{W}^{\rm ex})$. The black straight-line is the FT (\ref{equ::FT of ex work-log}). 
    Circles, triangles, and squares are respectively data from panels (a), (b), (c), whereas numbers are durations of processes.  Inset: The fitting slopes and error bars for each process.
  }
  \label{fig::240226-gamma1_gd003_ft_ex_combined_TR}
\end{figure}

Finally we verify Eq.~(\ref{equ::FT of ex work}), which may be rewritten as in Eq.~(\ref{equ::FT of ex work-log}). We simulate the same processes as shown in Table \ref{table::1028-ft_wex_combined_TR}, and compute the distributions of excess work.  We then do the same for the adjoint backward processes. All work distributions are displayed in Fig.~\ref{fig::240226-gamma1_gd003_ft_ex_combined_TR} (a), (b), and (c). Finally, we use these distributions to verify the FT  (\ref{equ::FT of ex work-log}).  As shown in Fig.~\ref{fig::240226-gamma1_gd003_ft_ex_combined_TR} (d), all data collapse to the black straight-line as predicted by our theory.  

\end{document}